\documentclass[english]{iopart}

\usepackage{graphicx}
\usepackage{url}
\usepackage{iopams} 

\expandafter\let\csname equation*\endcsname\relax
\expandafter\let\csname endequation*\endcsname\relax

\usepackage{amssymb}
\usepackage[latin1]{inputenc}
\usepackage{amsmath}
\usepackage{epsfig}
\newcounter{fig}
\begin{document}

\title[Singularities of lattice models ]
{\Large Holonomic functions of several complex variables and
singularities of anisotropic Ising $\,n$-fold integrals}
\vskip .3cm

\author{S. Boukraa$||$,  S. Hassani$^\S$, 
J-M. Maillard$^\pounds$}
\address{$||$  \ LPTHIRM and D\'epartement d'A{\'e}ronautique,
 Universit\'e de Blida, Algeria}
\address{\S  Centre de Recherche Nucl\'eaire d'Alger, 
2 Bd. Frantz Fanon, B.P. 399, 16000 Alger, Algeria}
\address{$^\pounds$ LPTMC, UMR 7600 CNRS, 
Universit\'e de Paris 6, Tour 23,
 5\`eme \'etage, case 121, 
 4 Place Jussieu, 75252 Paris Cedex 05, France} 

\vskip .3cm

{\em Dedicated to Fa Yuen Wu on the occasion of his 80th birthday.}

\vskip .1cm

\begin{abstract} 
Focusing on examples associated with holonomic functions,
we try to bring new ideas on how to look at phase transitions, 
for which the critical manifolds are not points but  curves depending
on a spectral variable, or, even, fill higher dimensional submanifolds.
Lattice statistical mechanics, often provides a natural (holonomic) framework 
to perform singularity analysis with several complex variables 
that would, in the most general mathematical 
framework, be too complex, or simply could not be defined. In a learn-by-example 
approach, considering several Picard-Fuchs systems 
of two-variables ``above'' Calabi-Yau ODEs, associated with double 
hypergeometric series, 
we show that D-finite (holonomic) functions are actually a good framework 
for actually finding properly the singular manifolds. The 
singular manifolds are found to be genus-zero curves. We, then, analyse 
the singular algebraic varieties of 
quite important holonomic functions of 
lattice statistical mechanics, the $\,n$-fold integrals $\, \chi^{(n)}$,
corresponding to the $\, n$-particle decomposition of the
magnetic susceptibility of the anisotropic 
square Ising model. In this anisotropic case, we revisit a set of 
so-called ``Nickelian singularities'' that turns out to be a two-parameter family 
of elliptic curves. We then find a first set of non-Nickelian
singularities for $\, \chi^{(3)}$ and $\, \chi^{(4)}$, that also turns out to 
be rational or ellipic curves. We underline the fact that these
singular curves depend on the anisotropy of the Ising model, 
or, equivalently, that they depend on the spectral parameter
of model. This has important consequences on the physical nature 
of the anisotropic $\, \chi^{(n)}$'s which appear to be highly  
composite objects. We address, from a birational viewpoint, the 
emergence of families of 
elliptic curves, and of Calabi-Yau manifolds on such problems.
We also address the question of the singularities of non-holonomic
functions with a discussion on
the accumulation of these singular curves for the non-holonomic 
anisotropic full susceptibility  $\, \chi$. 

\end{abstract}

\vskip .1cm


\noindent {\bf AMS Classification scheme numbers}: 34M55, 
47E05, 81Qxx, 32G34, 34Lxx, 34Mxx, 14Kxx 
\vskip .2cm


\section{Introduction}
\label{introduc}

Singularities are known to play a crucial role 
in physics (particle physics~\cite{Eden},
Landau singularities~\cite{Landau,Landau2}, 
critical phenomena theory, renormalization group, 
dynamical systems).
They are the ``backbone'' of many physical phenomena, in the same way 
cohomology can be introduced in mathematics as a ``skeleton'' describing
the most fundamental part of so many mathematical problems\footnote[3]{And not
surprisingly, cohomology is naturally introduced in the 
singularity theory~\cite{Bedford}.}. 

Seeking for the singular points, and/or critical manifolds of models
in lattice statistical mechanics is a necessary preliminary 
step towards any serious study of the lattice models. If the model 
is Yang-Baxter integrable there is a canonical parametrization of the
model in algebraic varieties~\cite{Automorphism}, and the critical manifolds
will also be algebraic varieties. If one does not expect 
 the model to be ``integrable'' (or even that the integrability of the  
model requires too much work to be performed), finding the singular manifolds
of the model is an attempt to obtain, at least, one exact result 
for the model. Recalling the standard-scalar Potts model~\cite{Wureview,WuPotts}, 
it is worth keeping in mind that its singular
 manifolds (corresponding to second order phase
transitions or first order phase transitions) are selected codimension-one 
algebraic varieties where the model is actually Yang-Baxter integrable.
The crucial role played by the (standard-scalar) Potts model 
in the theory of critical phenomena, 
is probably at the origin of  some ``conformal theory'' mainstream prejudice
identifying criticality with integrability for two-dimensional models. 

A large number of papers~\cite{challenge,challenge2,challenge3} have tried 
(under the assumption of a unique phase transition)
to obtain critical, and more generally singular\footnote[2]{If the wording ``critical'' 
still corresponds to singular in mathematics, it tends to be associated with
second order phase transitions exclusively. The singular condition for the standard-scalar  
$\, q$-state Potts model corresponds to second order phase transitions for $\, q < 4$
and first order transitions for $\, q >4$.}, manifolds of lattice models 
as algebraic varieties preserved by some (Kramers-Wannier-like) duality, 
thus providing, at least, one exact (algebraic) result for the model, 
and, hopefully, algebraic subvarieties candidates for Yang-Baxter integrability
of the models. The relation between singular manifolds of lattice statistical
models and integrability is, in fact, 
{\em much more complex}. Along this line it is worth recalling two examples.

A first example is the sixteen vertex model which is, generically, {\em not Yang-Baxter 
integrable}, but is such that the birational symmetries of the $CP_{15}$
parameter space of the model are {\em actually integrable}\footnote[1]{We have called
such models ``Quasi-integrable'': they are {\em not} Yang-Baxter integrable
but the birational symmetries of their parameter space 
correspond to {\em integrable mappings}~\cite{sixteen}.}, thus yielding a canonical
parametrization\footnote[8]{A foliation of $\, CP_{15}$ in 
elliptic curves.} of the model in terms of {\em elliptic curves}~\cite{sixteen}.
This parametrization
gives natural candidates for the singular manifolds of the model, namely the
vanishing condition of the corresponding $\, j$-invariant (which is actually 
the vanishing condition of a homogeneous polynomial of degree 24
 in the sixteen
 homogeneous parameters of the model, the polynomial being the sum 
a very large\footnote[9]{In~\cite{sixteen}
this  polynomial of degree 24 in 16 unknowns 
 is seen as 
the double discriminant of a biquadratic. It is nothing but
a hyperdeterminant~\cite{hyperdet,hyper} 
(Sch\"afli's hyperdeterminant~\cite{Schafli} 
of format 2 x 2 x 2 x 2).
It has 2894276 terms.} number of monomials~\cite{sixteen}).
 This codimension-one algebraic variety 
is, probably, not Yang-Baxter integrable.

A second example is the triangular $\, q$-state Potts model with 
three-spin interactions on the up-pointing 
triangles~\cite{inversionWu,inversionRollet} for which the critical manifold
has been obtained as a simple codimension-one algebraic variety~\cite{WuZia}.
This  codimension-one algebraic variety is a remarkable selected one: 
it is preserved by a ``huge'' 
set of birational transformations~\cite{Coxeter,Coxeter2}.
Recalling the previous ``conformal theory'' prejudice on standard-scalar 
$\ q$-state Potts models, it is worth mentioning that, {\em even 
restricted to this singular codimension-one} 
algebraic variety, the model {\em is not\footnote[5]{It is not 
Yang-Baxter integrable in the natural embedding of the model (namely
a parameter space made of the three (anisotropic) 
nearest neighbour edge interactions and the three-spin interaction 
on the up-pointing triangle).
Of course, it is always conceivable, that, upon increasing the parameter space, the 
selected critical algebraic subvariety becomes embedded in a Yang-Baxter family.
However the hyperbolic character~\cite{inversionWu,inversionRollet,Coxeter,Coxeter2} 
of the set of birational automorphisms of this 
algebraic subvariety seems to exclude an {\em abelian variety} for the 
larger (integrable) variety. Furthermore, random matrix analysis 
also seemed to exclude an integrability of this subvariety.} Yang-Baxter integrable}.

People working on lattice statistical mechanics (or condensed matter theory)
have some (lex parsimoniae\footnote[2]{Ockham's razor.}) 
simplicity prejudice that there exists a concept
of ``singularities of a model'', the singularities of the partition function
being, ``of course'', the same as the singularities of the full susceptibility. 
Furthermore, they also have another simplicity prejudice, namely that 
singularity manifolds are simple sets, like points, 
 (self-dual) straight lines, smooth codimension-one manifolds,
the maximum complexity being encountered with the phase diagram of 
the Ashkin-Teller model~\cite{AT}, with the emergence of tricritical 
points~\cite{tricritWu,Meyer}, forgetting less common (and more sophisticated
or involved) critical behaviors like, the Kosterlitz-Thouless 
transition~\cite{KT}, the massless phase in the 
classical XY model or in $\, Z_N$
 models (see for instance~\cite{Cardy,Alcaraz}), or the 
massless phase in the 
 three-state superintegrable 
chiral Potts model~\cite{Albertini} or in the XXZ 
quantum chain~\cite{Albertini,Albertini2,Maillet}, 
the Griffiths-McCoy singularities~\cite{McCoyBook,McCoyGriff}
 in random systems and the much more complex 
phase diagrams of commensurate-incommensurate models~\cite{Selke,Selke2,chiral,chiral2}.
This Ockham's razor's simplicity prejudice is clearly not shared
by people working on singularity theory in algebraic geometry, 
and discrete dynamical systems~\cite{Bedford,PC,Noetherian}
(see also Arnold's viewpoint on singularity theory
and catastrophe theory~\cite{Arnold}). 

In fact, singular manifolds in lattice statistical mechanics
(or condensed matter theory) have no reason to be simple codimension-one
sets (or even stratified spaces). For lattice models of statistical mechanics,
 where the parameter space corresponds to
{\em several} (complex) variables, there is a gap between a physicist's viewpoint that 
roughly amounts to seeing singular manifolds as simple mutatis-mutandis 
generalizations of singularities of 
one complex variable, conjecturing singular manifolds
as algebraic varieties~\cite{challenge,challenge2,challenge3}, and the 
mathematician's viewpoint that is reluctant to introduce the concept of 
singular manifolds for functions of several complex variables 
(it is not clear that the functions one studies are even defined 
in a Zariski space).

Singular manifolds can be well-defined in a framework that
is, in fact, quite natural, and 
{\em emerges quite often in theoretical physics}, namely 
the {\em holonomic functions}~\cite{Kawai} corresponding to $\, n$-fold integrals 
of a holonomic integrand (most of the time, in theoretical physics, the integrand 
is simply rational or algebraic). 
In  Sato's $\, {\cal D}$-module theory~\cite{Dmodule}, a holonomic system 
is a highly {\em over-determined} system, 
such that the solutions locally form a vector space of {\em finite dimension}
 (instead of the expected dependence on some arbitrary 
functions). Furthermore, holonomic functions naturally 
correspond to systems with {\em fixed regular 
singularities}. It is crucial to avoid movable singularities.
For {\em non-holonomic functions}, only the ones that 
{\em can be decomposed as an infinite
sum of holonomic functions} (like $\, \chi$,  the full susceptibility 
of the square\footnote[1]{We have similar decompositions as an infinite
sum of $\, n$-fold integrals for the full susceptibility
of the triangular or honeycomb Ising models for which 
dramatic extensions of their series expansion have 
been obtained recently~\cite{Chan}.} Ising model~\cite{wu-mc-tr-ba-76}) 
give some hope for interesting and/or rigorous studies of their singularities. 

\vskip.1cm

For one complex variable, the  holonomic 
(or D-finite~\cite{Lipshitz,Dfinitude}) functions are solutions 
of linear ODEs with polynomial coefficients in the 
complex variable. The (regular) singularities can be seen 
immediately as solutions of 
 the head polynomial coefficient of the linear ODE, up to
 apparent singularities~\cite{Khi3}.
If one takes a representation of the linear ODE 
as a linear differential system, one gets rid of the
apparent singularities, and one also sees, quite immediately,
 the singularities in such systems. 
More generally, for holonomic functions of several complex variables,
one can define, and see, quite clearly, the singular manifolds 
of the corresponding systems of PDEs. 
In a learn-by-example approach, we will show how one can find, and see, these 
singular algebraic varieties. 

\vskip .1cm 

The paper is organized as follows. After briefly recalling 
the framework of the isotropic $\, \chi^{(n)}$'s, 
we will first study various examples of Picard-Fuchs systems 
of two variables associated with hypergeometric series, 
and generalizing some known Calabi-Yau ODEs~\cite{Batyrev}. We will show 
how the singular manifolds can be obtained from the
holonomic systems, and from simpler asymptotic calculations. We will 
then obtain singular manifolds for quite important holonomic functions of 
lattice statistical mechanics, the $\,n$-fold integrals
$\, \chi^{(n)}$'s (corresponding
to the decomposition of the magnetic susceptibility of the anisotropic 
square Ising model~\cite{wu-mc-tr-ba-76}), describing 
a set of (so-called) ``Nickelian'' 
singularities, and then getting, 
from a ``Landau singularity~\cite{Eden,Landau} approach'',  
 a first set of  other (non-Nickelian) singularities.  We will underline 
the dependence of the singularity manifolds in the {\em anisotropy} 
of the Ising model.  This has important consequences 
for understanding the mathematical, as well as the physical,
 nature of the anisotropic $\, \chi^{(n)}$'s. The question 
of the accumulation of these
singular manifolds for the anisotropic full susceptibility  
$\, \chi$, will be discussed. We will finally comment on
the emergence of families of elliptic curves for the singularity manifolds, 
and the (birational) reason of the {\em occurrence of 
Calabi-Yau manifolds} on such problems.  

\vskip .1cm 

\section{Holonomic functions of one complex variable: the 
$\, \chi^{(n)}$'s for the isotropic Ising model}
\label{warm up}

Let us start with the simplest holonomic, or 
D-finite~\cite{Lipshitz}, functions, namely the 
holonomic functions of {\em one} complex variable, by recalling important holonomic 
functions of lattice statistical mechanics, 
the $\,n$-fold integrals $\, \chi^{(n)}$ of the isotropic 
square lattice Ising model~\cite{Khi3,Khi4,Experimental}.
These $\, n$-fold integrals correspond to the decomposition 
of the full susceptibility of the model
as an {\em infinite sum}~\cite{wu-mc-tr-ba-76} of 
the $\, n$-particle contributions
$\, \chi^{(n)}$. The 
singularities of these $\, \chi^{(n)}$'s have been completely described 
and can be seen to be a very rich and complex set of points~\cite{Landau2,SingNfold}.
In particular, one finds, in some well-suited variable $\, k$, which is the modulus of
the elliptic function parametrizing the two-dimensional
 Ising model, that the unit circle $\, |k |\, = \, 1\, $ will be
a {\em natural boundary} for the full susceptibility $\, \chi$
 of the Ising model~\cite{SingNfold}.
The singularities of the $\, \chi^{(n)}$'s accumulate on the unit circle. 
This is the reason why we have this unit circle 
{\em natural boundary}~\cite{Experimental,SingNfold,ongp,nic1,nic2}
 for the full magnetic susceptibility $\, \chi$.
Singularities {\em also} accumulate {\em inside} the unit circle 
(see Figures 1, 2, 3, 4 of~\cite{SingNfold}), probably 
becoming an {\em infinite set of points 
 dense in the open disk} $\, |k| \, < \, 1$. They 
also accumulate outside the unit $\, k$-circle 
$\, |k| \, > \, 1$, probably becoming another
 infinite set of points {\em also dense outside 
the unit circle}  $\, |k| \, > \, 1$. This accumulation 
of singular points of the linear ODEs
of the $\, \chi^{(n)}$'s is thus (probably) dense
 in the {\em whole $\, k$-complex plane}. 
In other words, we do have an infinite set of singularities {\em dense in 
the whole $\, k$-complex plane}. This seems 
to confirm the mathematician's reluctance
to consider singular manifolds of functions of several complex variables
that are not holonomic: even in the very simple case of {\em one} complex variable,
we already seem to encounter serious troubles. The full susceptibility  $\, \chi$,
which is an infinite sum~\cite{wu-mc-tr-ba-76} of these $\, \chi^{(n)}$'s, does 
not even seem to be defined in a Zariski space. Recalling these 
results~\cite{SingNfold}, the common wisdom identifying the singularities
of the partition function and the singularities
of the full susceptibility is no longer obvious.

There is, however, an important subtlety here: these singularities are {\em singularities
of the linear ODEs} of the $\, \chi^{(n)}$'s, but not of the 
(series expansions of the) $\, \chi^{(n)}$'s given by holonomic $\,n$-fold 
integrals. When one considers
 the $k$-series expansions for the
$\, \chi^{(n)}$'s, one finds out that the singularities inside the unit circle 
in the open disk $\, |k| \, < \, 1$, are {\em not singularities
 of these series}~\cite{SingNfold}. This is a quite non-trivial result.
This is also the case for the  $k$-series expansion 
for the  full susceptibility $\, \chi$ which is the infinite sum of the $\, \chi^{(n)}$'s. 
For the full susceptibility $\, \chi$, the accumulation
 of $\, \chi^{(n)}$'s singularities
on the unit circle makes this unit circle
 a {\em natural boundary}~\cite{SingNfold}. Switching 
from high-temperature
series expansions to low-temperature series, we have a similar
 result for $\, |k| \, > \, 1$. We thus have a quite drastic difference between 
the singularities of the $\,n$-fold integrals $\, \chi^{(n)}$,  which are solutions
of linear ODEs (they are D-finite or holonomic, see below),
and the full susceptibility  $\, \chi$ which is {\em not} solution
of a linear ODE (it is {\em not} holonomic). 

Before generalizing to several complex variables 
with the case of the $\, \chi^{(n)}$'s
for the {\em anisotropic} square Ising model with two complex variables, 
let us consider, in a learn-by-example 
approach, simple Picard-Fuchs systems associated with hypergeometric series
of {\em two} complex variables.  

\section{A first simple Picard-Fuchs system with two variables}
\label{picard}

Let us consider the double hypergeometric 
series, {\em symmetric in $\, x$ and $\, y$}
\begin{eqnarray}
\label{H2}
\hspace{-0.8in}&&H_0(x, \, y) 
\,  \, \, = \, \, \,  \, \, 
 \sum_{n=0}^\infty \,  \,  \sum_{m=0}^\infty \,  \,
 {{ (3m+3n)! } \over { n!^3 \, \, m!^3 }}
\cdot x^n \cdot y^m  \\
\hspace{-0.8in}&&\quad \quad\,\,  \, = \, \, \, \,  \,\, 
 \sum_{n=0}^\infty \,  \, {{(3\, n)!} \over {n!^3}} \, \cdot \,
 _3F_2\Bigl([n\, +1, \,n\, +{{1} \over {3}}, \, n\, +{{2} \over {3}}], 
\, [1, \, 1]; \, 27 \, y   \Bigr)
 \cdot \, x^n
\end{eqnarray}
\begin{eqnarray}\hspace{-0.8in}&&\quad \quad\,\,  \, = \, \, \, \,  \,\, 
1 \,\,\,   +6 \cdot (x+ y)\, \, \,  \,\, 
+(90 \cdot (x^2\, +y^2) +720 \cdot x \, y)  \, 
 \nonumber \\
\hspace{-0.8in}&&\quad\quad \quad \quad  \quad 
+(1680 \cdot (x^3+y^3) \, +45360 \cdot x \, y \, \cdot (x\, +y))
 \, \,\, +(34650 \cdot (x^4\, +y^4)
 \nonumber \\
\hspace{-0.8in}&&\quad \quad \quad \quad  \quad 
 + 2217600 \cdot x \, y \, \cdot (x^2\, +y^2) +7484400 \cdot x^2 \, y^2) 
\,\, \,  + \, \cdots    \nonumber  
\end{eqnarray}
This series reduces, when $\, y \, = \, x$, to 
\begin{eqnarray}
\label{solBat1}
\hspace{-0.9in}&& \sum_{n=0}^\infty \, 
\Bigl[{\frac{(3 n)!}{(n!)^3}} \, \sum_{k=0}^n {n \choose k}^3\Bigr] \cdot \, x^n 
\,\, \, \, = \,\, \, \,\,\,\, 
1 \,\,\, +\, 12 \cdot x \,\,\, +\, 900 \cdot x^2 \,\, +\, 94080 \cdot x^3 \,\, 
\nonumber \\
\hspace{-0.9in}&& \quad \quad \quad \quad 
\, +\, 11988900 \cdot x^4\,
\,  +1704214512 \cdot x^5  \, +260453217024 \cdot x^6 \, \,
\, + \,\, \cdots,  
\end{eqnarray}
which is the solution analytic at $\, x \, = \, \, 0$ 
of the order-four {\em Calabi-Yau operator} $\, \Omega$
introduced by Batyrev and van Straten (section 7.1 of~\cite{Batyrev}, see
also the ODE number 15 in~\cite{TablesCalabi}) 
\begin{eqnarray}
\label{Batyrev1}
\hspace{-0.6in}&&  \Omega \, \,  \, = \, \, \,  \, \, \,
 \theta^4 \,\, \,  -3 \, x \cdot 
(7\, \theta^2 \, +7 \,\theta \, +2) \cdot 
  (3\, \theta \, +\, 1) \cdot   (3\, \theta \, +\,2)
\, \nonumber \\
\hspace{-0.6in}&&\quad \quad \qquad \, \,\,  -72 \, x^2 \cdot  
   (3\, \theta \, +\, 5) \cdot   (3\, \theta \, +\,4) \cdot
   (3\, \theta \, +\, 2)   \cdot  (3\, \theta \, +\, 1),    \\
\hspace{-0.6in}&& \quad \quad  \quad \hbox{where:} \quad  \quad  \qquad
 \theta \,\,  = \, \, \, x \cdot {{d} \over {d x}}.
 \nonumber 
\end{eqnarray}

The double hypergeometric series (\ref{H2}) is the {\em unique} 
analytical (in $\, x$ and $\ y$)
solution of the Picard-Fuchs system corresponding to the
 two partial linear differential operators:
\begin{eqnarray}
\label{picard}
\hspace{-0.8in}&&\Omega_x \,\,  = \, \, \, \,  \,  \,
\theta_x^3 \,  \, \,\, \, - \, x \cdot (3 \, \theta_x \, + 3 \, \theta_y \, + \, 1) 
\cdot (3 \, \theta_x \, + 3 \, \theta_y \, + \, 2) \cdot 
(3 \, \theta_x \, + 3 \, \theta_y \, + \, 3),
 \nonumber \\
\hspace{-0.8in}&&\Omega_y \,\,  = \, \, \,  \,  \, \,
\theta_y^3 \, \,  \,\, \,
 - \, y \cdot (3 \, \theta_x \, + 3 \, \theta_y \, + \, 1) 
\cdot (3 \, \theta_x \, + 3 \, \theta_y \, + \, 2) \cdot 
(3 \, \theta_x \, + 3 \, \theta_y \, + \, 3), 
\end{eqnarray}
\begin{eqnarray}
\hspace{-0.8in}&& \quad  \quad  \quad \hbox{where:} \quad  \quad  \qquad 
\theta_x \,\,  = \, \, \, x \cdot {{\partial} \over {\partial x}}, 
\qquad \quad 
 \theta_y \,\,  = \, \, \,  y \cdot {{\partial} \over {\partial y}}.
 \nonumber
\end{eqnarray}

The other formal series solutions of (\ref{picard}), 
around $\,(x, \, y) \, = \, \,(0,0)$, have the form
\begin{eqnarray}
\label{r}
  H_0(x, \, y) \cdot \ln(x)^n \cdot \ln(y)^{m} \,+ \, \cdots  
\end{eqnarray}
where the maximum value reached by $n$ and $m$ is 2. They read for instance:
\begin{eqnarray}
\hspace{-0.9in}&&H_0(x, \, y) \cdot \ln(x) \, + \, H_1(x, \, y), 
\qquad \quad \quad H_0(x, \, y) \cdot \ln(y) \, + \, H_1(y, \, x), 
 \nonumber \\
\hspace{-0.9in}&&H_0(x, \, y) \cdot \ln(x) \cdot \ln(y) \, 
+ \, H_1(y, \, x) \cdot \ln(x) \, 
+ \, H_1(x, \, y) \cdot \ln(y)  \, + \, H_3(x, \, y), 
\quad \quad  \cdots 
 \nonumber 
\end{eqnarray}
It is crucial to note that the dimension of the 
space spanned by these formal series is  {\em finite}. 
In the case of the Picard-Fuchs system (\ref{picard}), the 
number of solutions (i.e. dimension) is nine.
These nine formal solutions are given in \ref{nineformal}.
The double series analytic in $\, x$ and $\, y$, $\, H_j(x, \, y)$ 
are either symmetric like $\, H_0(x, \, y)$, $\, H_3(x, \, y)$, 
or are not symmetric like $\, H_1(x, \, y)$.

Such holonomic systems are also called  
{\em D-finite}~\cite{Lipshitz,Dfinitude}, 
for that reason: remarkably, they have a {\em finite} number 
of independent solutions, in contrast 
with generic systems of PDEs that have, generically, an infinite number
of solutions. Systems of PDEs can also have no solution at all.
Generically the compatibility of the two operators $\, \Omega_x$ and 
$\, \Omega_y$, requires some (slightly tedious) differential algebra 
calculations.

One can also see the system (\ref{picard}) 
as a (two-dimensional) recursion:
\begin{eqnarray}
\hspace{-0.8in}&&(n \,  \, +1)^3 \cdot \, c_{n +1, \, m} 
\, \, = \, \, \, \,  b(n, \, m)  \cdot \,   c_{n, \, m}, 
\nonumber \\
\hspace{-0.8in}&&(m \,  \, +1)^3 \cdot \, c_{n, \, m+1} 
\, \, = \, \, \, \,  b(n, \, m)   \cdot \, c_{n, \, m}, 
\qquad \qquad  \quad  \hbox{where:}
\nonumber \\
\label{defbnm}
\hspace{-0.8in}&&\qquad \quad  b(n, \, m) 
\, \, \, = \, \, \, \, \, (3\, (n +m) \, +\, 1) 
\cdot (3\, (n +m) \, +\, 2) \cdot
 (3\, (n +m) \, +\, 3),
\end{eqnarray}
 Here, the compatibility between 
the two partial differential operators
$\, \Omega_x$ and $\, \Omega_y$ is easier 
to see at this (double) recursion level. 
Introducing 
\begin{eqnarray}
\hspace{-0.7in}&&\alpha_1(n, \, m) \, = \, \, 
 {{ b(n, \, m)  } \over { (n \,  \, +1)^3}} \, = \, \, 
{{c_{n +1, \, m} } \over { c_{n, \, m} }}, 
\quad \quad \quad \alpha_2(n, \, m) \, = \, \, \,
 {{ b(n, \, m)  } \over { (m \,  \, +1)^3}}\, = \, \, 
{{ c_{n, \, m+1}} \over { c_{n, \, m} }}, 
\nonumber
\end{eqnarray}
we have the identity:
\begin{eqnarray}
\alpha_2(n, \, m)  \cdot \alpha_1(n, \, m\, +1)  
 \,\, \, \,= \, \, \, \,\,    \alpha_1(n, \, m) \cdot 
 \alpha_2(n\, +1, \, m),  
\end{eqnarray}
which, from a recursion viewpoint, actually corresponds
{\em to the compatibility between 
the two partial linear differential operators}
$\, \Omega_x$ and $\, \Omega_y$.

\vskip .2cm 

The discriminant of the two-parameter family of Calabi-Yau 3-folds
reads\footnote[2]{Note a misprint in Prop. 7.2.1 of~\cite{Batyrev}:
$ \, (x \, + \, y)$ must be changed into  $\, 3 \cdot \, (x \, + \, y)$ .
} (see Prop. 7.2.1 of~\cite{Batyrev}):
\begin{eqnarray}
\label{discrimCalab}
 (x \, + \, y)^3 \,\, \,
 -\, 3 \cdot \, (x^2 \, -7 \, x \, y \, + \, y^2) \,\,  \,
+\, 3 \cdot \, (x \, + \, y) \,\, \, \, -1, 
\end{eqnarray}
or, (without performing the  $\, (x, \, y) \, \rightarrow 
\, (\, x/27, \, \,y/27) \, $
rescaling mentioned in~\cite{Batyrev}):
\begin{eqnarray}
\label{discrim}
\hspace{-0.5in}\Delta \,\,  = \, \, \,\,
19683\, (x+y)^3 \,  \,\, -2187 \cdot \,(y^2+x^2-7\, x\, y)\, \, \,
 +81 \cdot \, (x+y)\, \, \, -1. 
\end{eqnarray}
This expression can easily be obtained as the resultant~\cite{hyperdet} 
in $\,A$ (or equivalently in $\, B$)
of the two (very simple) homogeneous binary cubics~\cite{Batyrev}:
\begin{eqnarray}
\label{cubic}
\hspace{-0.5in} 27\, x\cdot \, (A+B)^3\,  -A^3 \, \, = \, \, \, 0, 
\qquad \quad 
27\, y\cdot \, (A+B)^3\,  -B^3 \, \, = \, \, \, 0.
\end{eqnarray}

\vskip .1cm 

\subsection{Singular manifolds} 

\label{picardsing}
\vskip .1cm 

What are the singularities of the double hypergeometric series like (\ref{H2}),
and how do they compare with the singularities of the Picard-Fuchs system 
(\ref{picard}), assuming that the notion of  singularities 
of such PDEs systems is well-defined ?

From a mathematical viewpoint, when introducing some ``canonical'' system, equivalent 
to the Picard-Fuchs system, one should ``in principle'' be able to see 
the singularities as simple poles of this equivalent system. Unfortunately,
 to our knowledge, the implementation of such procedure is available
as formal calculation tools is still 
in development~\cite{Cluzeau} (see also~\cite{PDE,Barkatou}).

A physicist's down-to-earth approach amounts to reducing the
double hypergeometric series, like (\ref{H2}), to series in one (complex) variable
imposing some relation between $\, x$ and $\, y$, 
compatible with the $\,(x, \, y) \, = \, \, (0, \, 0)$
origin of the double series. Imposing, for example, $\, y \, = \,\, c \, x\, $ 
($c\, = 2, \, 3, ...$), or 
 $\, y \, = \,\, c \, x^2$, one gets a series in one (complex) variable
$\, x$ and, then, in the second step, finds the corresponding linear ODE annihilating
this series. The head polynomial of the corresponding linear differential
operator gives (after getting rid of the apparent singularities) the
singularities of these linear differential operators.
An ``accumulation'' of such results enables to see that the singularities
are always on the (genus-zero) algebraic curve $\, {\cal S}(x, \, y) \, = \, \, 0$,  
where 
\begin{eqnarray}
\label{cand}
\hspace{-0.7in} 
{\cal S}(x, \, y) 
\,  \, = \, \, \,  \, \, \,  \, 
 3^9 \cdot \, (x+y)^3 \, \,  \,  -\,3^7 \cdot \,(y^2+x^2-7\, x\, y)\, \, \, 
  +3^4 \cdot \, (x+y) \, \,  \, -1,  
\end{eqnarray}
which is {\em nothing but  the discriminant} (\ref{discrim})  {\em of 
 the two-parameters family of Calabi-Yau 3-folds
previously mentioned}~\cite{Batyrev}. Remarkably, but not surprisingly,  the
{\em singular variety has an interpretation as a fundamental 
projective invariant}~\cite{hyperdet}.

\vskip .1cm 

The (genus-zero) singular curve (\ref{cand}) can be
parametrized by
\begin{eqnarray}
\label{22}
\hspace{-0.2in}x \, \, = \, \, \,  \Bigl({{ 1} \over {6 }} \, + \, \, u\Bigr)^3,
 \qquad  \quad 
y \, \, = \, \, \,  \Bigl({{ 1} \over {6 }} \, - \, \, u\Bigr)^3.
\end{eqnarray}
or
\begin{eqnarray}
\hspace{-0.6in}x(u) \, \, = \, \, \,  
\Bigl( {{5 \,u \, + \, 7 } \over { 6 \cdot \, (1\, -u) }}  \Bigr)^3,
\qquad \, \, \,
y(u) \, \, = \, \, \,  
\Bigl(  {{ 7 \,u \, + \, 5 } \over { 6 \cdot \, (u\, -1) }}  \Bigr)^3
 \, \, = \, \, \, x\Bigl({{1} \over {u}} \Bigr), 
\end{eqnarray}
where the Atkin-Lehner-like involution $\, \, u \, \, \leftrightarrow  \,\, 1/u \, \, $ 
could suggest a modular curve interpretation of (\ref{cand}).

The accumulation of calculations is quite tedious compared to the simplicity
of the final result (\ref{cand}). It is far from obvious that (\ref{cand})
is the singularity manifold of the double series (\ref{H2}), or 
the singularity manifold  of the Picard-Fuchs system (\ref{picard}). Let us 
find a Picard-Fuchs system
for which it will become crystal clear
that (\ref{cand}) is actually the singularity manifold of the system. 

\subsection{Other representations as PDE systems}
\label{otherpicard}

In fact, the Picard-Fuchs partial differential system (\ref{picard})
can be recast into a system of two differential equations, each one 
being a linear ODE on {\em only one}
 variable. We consider\footnote[1]{For our purpose, we did not 
 use the Groebner basis approach (use the pdsolve command 
on the system of equations obtained from the 
Rosenfeld-Groebner command in Maple).}
a linear combination of $\, \Omega_x$, $\,\Omega_y$
and their derivatives, and cancel the coefficients 
in front of the undesired derivatives. We obtain the following form
\begin{eqnarray}
\label{form}
\hspace{-0.5in}&&{\tilde \Omega}_x \,\, = \,\,\,\,
 \sum_{n=0}^{9} \, P_n(x, \, y) \cdot \, D_x^n, 
\qquad \qquad {\tilde \Omega}_y \,\, = \,\,\,\,
\sum_{n=0}^{9} \,  Q_n(x, \, y) \cdot \, D_y^n, 
\end{eqnarray}
\begin{eqnarray}
\hspace{-0.5in}&& \qquad \hbox{where:}  \qquad \qquad 
 D_x \,\, = \,\,\, {{\partial} \over {\partial x}},
 \quad  \quad  \quad 
 D_y \,\, = \,\,\, {{\partial} \over {\partial y}}, 
\nonumber 
\end{eqnarray}
where $\, P_n(x, \, y)$ and $\, Q_n(x, \, y)$ are 
 polynomials of the two variables $\, x$ and $\, y$.
The partial differential operator $\, {\tilde \Omega}_x$ can be seen as a 
linear differential operator in $\, x$ depending on a parameter $\, y$
(and similarly $\, {\tilde \Omega}_y$ as a linear differential operator 
in $\, y$ depending on a parameter $\, x$). The polynomials $\, P_n(x, \, y)$ 
appearing in $\, {\tilde \Omega}_x$ 
will not be given here. For $\, P_9(x, \, y)$ the monomial of highest degree
 in $\, x$ and $\, y$ is $\,\, x^{15} \, y^9$ (see (\ref{appare}) and 
(\ref{calP9}) in \ref{L6}), 
and, for $\, P_8(x, \, y), \, \cdots,  \, P_0(x, \, y)$, 
it reads, respectively, 
$\,\, x^{14} \, y^9, \, \,  x^{13} \, y^9,  \,\,x^{12} \, y^9, \,\,x^{11} \, y^9$, 
$\,  x^{10} \, y^9, \, \, x^{9} \, y^9$, 
$\,  \,  x^{8} \, y^8, \,  \,x^{7} \, y^7, \,  \,x^{6} \, y^6$.    

\vskip .1cm

There is a ``price to pay'' to recast the Picard-Fuchs partial linear
differential system (\ref{picard})
into a system like (\ref{form}). The partial  linear differential 
operators $\, {\tilde \Omega}_x$ 
and $\, {\tilde \Omega}_y$ are {\em much more involved}
 than operators $\, \Omega_x$ and  $\, \Omega_y$ in (\ref{picard}),
and of higher order in $\, D_x$ or $\, D_y$. The
 operator $\, {\tilde \Omega}_x$ 
(resp. $\, {\tilde \Omega}_y$) is of {\em order nine} with respect 
to  $\, D_x$ (resp. $\, D_y$),
in agreement with the previously mentioned finite set (\ref{list1}) 
of {\em nine formal series solutions} of the  
Picard-Fuchs D-finite system (\ref{picard}). We have checked that 
these nine formal solutions (\ref{list1}) are indeed solutions of 
$\, {\tilde \Omega}_x$ (resp. $\, {\tilde \Omega}_y$).

As a consequence of the exact symmetry 
interchange $\,\, x \, \leftrightarrow \, y \, $ of (\ref{H2}), 
the partial differential operator $\, {\tilde \Omega}_y$ is nothing
 but operator $\, {\tilde \Omega}_x$,
where $\, x$ and $\, y$ are permuted. Not surprisingly, the head polynomials
 in (\ref{form})
have the form 
\begin{eqnarray}
\label{appare}
\hspace{-0.9in}\, \, P_9(x, \, y) \,  \, = \, \, \, 
x^6 \cdot \, {\cal P}_9(x, \, y) \cdot \, {\cal S}(x, \, y), 
\quad \, \, \, \, \, 
Q_9(x, \, y) \,  \, = \, \, \, 
 y^6 \cdot \,{\cal P}_9(y, \, x) \cdot \, {\cal S}(x, \, y),
\end{eqnarray}
where $\, {\cal P}_9(x, \, y)$ is a polynomial of $\, x$ and $\, y$,
 corresponding to the {\em apparent} singularities 
of the ($\, y$-dependent) linear differential 
operator $\, {\tilde \Omega}_x$. The expression 
of  $\, {\cal P}_9(x, \, y)$ is given in \ref{L6}.

\vskip .2cm 

\subsection{Operator factorizations}
\label{opfacto}

\vskip .1cm 
One can actually go further in the analysis of these order-nine
operators. The order-nine partial  linear differential 
operator $\, {\tilde \Omega}_x$, in fact, factorizes
in three order-one operators, and an order-six operator: 
\begin{eqnarray}
\label{factoL9L6}
\hspace{-0.6in}{\tilde \Omega}_x \,  \, \, = \, \, \, \,  \,
\Bigl(D_x \, - \, \, {{\partial \ln(\tilde{r}_1(x, \, y))
 } \over {\partial x}}  \Bigr) 
\cdot \, 
\Bigl( D_x \, - \, \, {{\partial \ln(\tilde{r}_2(x, \, y)) 
} \over {\partial x}}  \Bigr) 
\nonumber  \\
\hspace{-0.6in} \qquad  \qquad  \qquad  \qquad 
\times  \,  \,    \, 
\Bigl(D_x \,  - \, \, {{\partial  \ln(\tilde{r}_3(x, \, y))
 } \over {\partial x}}   \Bigr)
 \cdot \, \, L_6(x, \, y),
\end{eqnarray}
where  the order-six operator $\, L_6(x, \, y)$ reads
\begin{eqnarray}
\label{L6ordersix}
L_6(x, \, y) \,  \, \, = \, \, \,  \,\,  \,
{{1} \over {p_6(x, \, y)}}  \cdot \, \sum_{n=0}^6 \, p_n(x, \, y) \cdot D_x^n, 
\end{eqnarray}
and where $\, \tilde{r}_1(x,\, y)$, $\, \tilde{r}_2(x,\, y)$ 
and $\, \tilde{r}_3(x,\, y)$ are  rationals functions of $\, x$ and $\, y$,
while $\,p_6(x, \, y)$ has simple factorizations: 
\begin{eqnarray}
\label{qn}
\hspace{-0.8in}&&\tilde{r}_1(x,\, y) \,  = \, \, \, 
{{ {\cal P}_9(x, \, y) } \over { x^6 \cdot \, {\cal S}(x, \, y) \cdot \, q_1   }}, 
\quad      \qquad 
\, \, \, \, \, \,   \tilde{r}_2(x,\, y) \,  = \, \, {{q_1} \over 
{  x^5 \cdot \,  {\cal S}(x, \, y)\cdot \,  q_2 }}, 
\, \, \, \, 
\nonumber \\
\hspace{-0.8in}&&\tilde{r}_3(x,\, y) \,   = \, \, {{ q_2 } \over {
 x^4 \cdot \,  {\cal S}(x, \, y)\cdot \,  {\cal P}_6(x, \, y) }},  
\, \, \, \, \, \,  \, \, 
p_6(x, \, y) \,  = \, \, 
 x^4 \cdot \,  {\cal S}(x, \, y)\cdot \, {\cal P}_6(x, \, y), 
\end{eqnarray}
where  $\, {\cal P}_9(x, \, y)$, $\, {\cal P}_6(x, \, y)$, 
$\, q_1$, $\, q_2$,
are polynomials of $\, x$ and $\,y$
 given in \ref{L6}. 
Not surprisingly the $(x, \, y)$-asymmetric polynomials
$\, {\cal P}_6(x, \, y)$ and 
$\, {\cal P}_9(x, \, y)$ correspond respectively 
to apparent singularities of
the order-six and order-nine operators
 $\, L_6(x, \, y)$ and $\, {\tilde \Omega}_x$. The polynomials 
$\, p_n(x, \, y)$ appearing in $\,L_6(x, \, y)$ 
will not be given here. For $\, p_6(x, \, y)$ the monomial of highest degree
 in $\, x$ and $\, y$ is $\,\, x^{13} \, y^9$ (see (\ref{qn}) and 
(\ref{calP6}) in \ref{L6}), and, for
 $\, p_5(x, \, y), \, \cdots,  \, p_0(x, \, y)$, it reads, respectively, 
$\,\, x^{13} \, y^9, \, \,  x^{12} \, y^9,  \,\,x^{11} \, y^9$, 
$\,  x^{10} \, y^9, \, \, x^{9} \, y^9$, 
$\,  \,  x^{8} \, y^8, \,  \,x^{7} \, y^7$.    

Do note that the critical exponents of this order-six operator  $\, L_6(x, \, y)$ 
are {\em independent of} $\, y$. For instance at $ \, x \, = \, \, 0$
the indicial polynomial reads $\, P(r) \, = \, \, r^3 \cdot (r-1)^3$. 
More remarkably, {\em on the singular variety} 
 $\, {\cal S}(x, \, y) \, = \, \, 0$, the critical 
exponents of  $\, L_6(x, \, y)$ are
{\em also independent of} $\, y$. 
The indicial polynomial, at $\, {\cal S}(x, \, y) \, = \, \, 0$,
 reads 
$\, P(r) \, = \, \, r \cdot (r-1)^2 \cdot (r-2) \cdot (r-3) \cdot (r-4)$. 
The singular behaviour at $\, {\cal S}(x, \, y) \, = \, \, 0$ 
is thus logarithmic.
The wronskians of this order-six linear differential operator 
 $\, L_6(x, \, y)$, and of the order-nine  operator $\, {\tilde \Omega}_x$
 are rational functions of $\, x$ and $\, y$, which read
respectively:
\begin{eqnarray}
\label{wronski}
\hspace{-0.7in}W\Bigl(L_6(x, \, y)\Bigr) \,\, \, = \, \, \,\,  \,  \, 
{{ {\cal P}_6(x, \, y) } \over {x^{12} \cdot \, {\cal S}(x, \, y)^4}},
\qquad W\Bigl({\tilde \Omega}_x\Bigr) \,\, \, = \, \, \,\,  \,  \,  
{{
{\cal P}_9(x, \, y) } \over {x^{27} \cdot \, {\cal S}(x, \, y)^7}}.
\end{eqnarray}

In fact, the operator  $\, L_6(x, \, y)$ is not only {\em Fuchsian} with 
 {\em rational exponents} and   {\em rational wronskian}, 
it is actually {\em globally nilpotent}
for any rational values of $\, y$. The $\, p$-curvature of this 
globally nilpotent order-six operator, is a nilpotent $\, 6 \times 6$
matrix which can be put into the following 
Jordan form\footnote[1]{Of characteristic polynomial 
$\, P(\lambda) \, = \, \, \lambda^6$ and 
of minimal  polynomial $\, P_m(\lambda) \, = \, \, \lambda^4$.},
 {\em not only for any  rational value of} $\, y$, but, actually,
 {\em for any}  $\, y$ {\em being an algebraic number}:
 \vskip .1cm 
\begin{eqnarray}
\hspace{-0.6in}{\cal C} \, \, \, = \, \, \, \, 
\left[ \begin{array}{cccccc} 
0&1&0&0&0&0\\ 
\noalign{\medskip}0&0&0&0&0&0\\ 
\noalign{\medskip}0&0&0&1&0&0\\ 
\noalign{\medskip}0&0&0&0&1&0\\ 
\noalign{\medskip}0&0&0&0&0&1\\ 
\noalign{\medskip}0&0&0&0&0&0
\end{array} \right], \qquad \hbox{where:} \qquad \,\, \quad
{\cal C}^4 \, \, \, = \, \, \, \, 0. 
\end{eqnarray}
\vskip .1cm 

Furthermore  $\, L_6(x, \, y)$ is such that {\em its exterior square 
is of order fourteen}  
($D_x^ {14} \, + \, \cdots $) instead of the order-fifteen 
one should expect generically for an order-six irreducible operator.
This remarkable property is related to the fact that  $\, L_6(x, \, y)$
is {\em homomorphic to its (formal) adjoint}, with an order-two 
intertwinner differential operator $\, I_2(x, \, y)$
\begin{eqnarray}
\label{Homoadjoint}
\hspace{-0.8in}&&L_6(x, \, y)   \cdot \, I_2(x, \, y) 
\, \, \, \, =  \, \, \, \,  
adjoint(I_2(x, \, y))  \cdot adjoint(L_6(x, \, y)), 
\qquad \quad \hbox{where:}  \nonumber \\
\hspace{-0.8in}&&\qquad I_2(x, \, y) \, \,\, =  \, \, \, \, 
3^6 \cdot \, {{27\,x +27\, y \, +2} \over { 
{\cal S}(x, \, y)}} \cdot \, D_x^2 \,\,  
+ \, R_1(x, y) \cdot \, D_x \,\,  +\, R_0(x, y), 
\end{eqnarray}
where $\, R_1(x, \, y)$ and $\, R_2(x, \, y)$ are rational functions 
of $\, x$ and $\, y$. 

One can check that the double $(x, \, y)$-symmetric 
series (\ref{H2}), solution of the 
order-nine  operator $\, {\tilde \Omega}_x$,
is, in fact, annihilated by the order-six linear differential operator 
$\, L_6(x, \, y)$ and, thus (by 
$\, x \, \leftrightarrow \, y \, $ symmetry)
by the other order-six operator
\begin{eqnarray}
L_6(y, \, x) \,  \, \, = \, \, \,  \,\,  \,
{{1} \over {p_6(y, \, x)}}  \cdot \, \sum_{n=0}^6 \, p_n(y, \, x) \cdot D_y^n. 
\end{eqnarray}

\vskip .1cm 

At this step, we should recall that our purpose is to get the singularities
of the system (\ref{picard}) and not to obtain an equivalent system 
for (\ref{picard}).
Generically, systems of linear PDEs 
{\em cannot be strictly recast}\footnote[2]{Non-holonomic systems 
cannot be recast into a form like (\ref{form}).
This is the case, for instance, of the system of linear operators
 $\, (\Omega_x, \,\Omega_y)  \, = \,\, (D_x^2, \, D_x  \, D_y)$, 
which has an {\em infinite number} of solutions, namely 
 $\, c \cdot \, x \, + \, f(y)$ where $\, f(y)$ is an 
{\em arbitrary function of} $\, y$.}
into a form like (\ref{form}), even for D-finite systems\footnote[3]{
For instance, the solutions of the D-finite system $\, (\Omega_x, \,\Omega_y)  $
 $\, = \,\, (D_x^2\, -y \, D_y^2, \, D_x  \, D_y)$ are solutions of 
the D-finite system  $\, ({\tilde \Omega}_x, \, {\tilde \Omega}_y)  $
$\, = \,\, (D_x^3, \, y  \, D_y^3 \, +\, D_y^2)$, but this last D-finite 
system has more solutions. One needs additional operators,
to have  a system equivalence.}.
The two order-six  operators  $\, L_6(x, \, y)$ and $\, L_6(y, \, x)$
form a PDE system that is {\em not equivalent} (in the sense of equivalence
of systems) to the Picard-Fuchs system (\ref{picard}).
However, and as far as the
double series $\, H_0(x,y)$ is concerned, the three systems $\, (\Omega_x, \,\Omega_y)$,
 $\, ({\tilde \Omega}_x, \, {\tilde \Omega}_y)$, or $\, (L_6(x, \, y), \,  L_6(y, \, x))$,
can alternatively be considered.

\vskip .1cm 

{\bf Remark:} Recovering the  Calabi-Yau order-four ODE (\ref{Batyrev1}) from 
the $\, y \, = \, x$ limit of the Picard-Fuchs system (\ref{picard}),
or (\ref{form}), is {\em not straightforward} (as one could naively imagine).
Within the (down-to-earth) approach which amounts, for instance, to 
restricting to the straight lines $\, y \, = \, c \cdot \, x$, where $\, c$ 
is a constant, and finding the linear differential operator in $\, x$, 
one obtains an order-six linear differential operator with 
coefficients that are  polynomials in $\, x$, as well as
 in the constant $\, c$. One can, then,
take the $\, c\, \rightarrow \, 1$ limit and actually recover
the  Calabi-Yau order-four ODE (\ref{Batyrev1}). 
These calculations are displayed in \ref{further2}.
The (genus-zero) singular curve (\ref{cand}) 
\begin{eqnarray}
\label{singsing}
\hspace{-0.5in}(1 \, -108 \cdot (x+y))  \cdot (2+27 \cdot (x+y))^2 \, 
\, \, +\, 3^9 \cdot (x-y)^2 \,\,\,\, = \,\,\,\, \, 0, 
\end{eqnarray}
 reduces, in the $\, y \, = \, \, x$ limit, to 
$\, (1\, -216 \, x) \cdot (1\, -27 \, x)^2 \,  =  \, \, 0$,
namely the singularities corresponding to the 
order-four Calabi-Yau  ODE (\ref{Batyrev1}).

\section{More Picard-Fuchs systems with two variables}
\label{picardmore}

Similar calculations can be performed with double hypergeometric series
generalizing the analytic solution of another Calabi-Yau order-four ODE
(see \ref{Kampe} below). One can perform exactly the same calculations 
mutatis mutandis. 

\subsection{More Picard-Fuchs system with two variables}
\label{picardmorebat2}
\vskip .1cm
 Let us, first, consider a two-variables Picard-Fuchs system  
``above'' another Calabi-Yau ODE~\cite{Batyrev} (see the ODE number 16
 in appendix A of~\cite{TablesCalabi}),  corresponding 
to the following $(x, \, y)$-symmetric series with {\em binomial} coefficients:
\begin{eqnarray}
\label{24}
\hspace{-0.9in}&&\sum_{n\, = \, 0}^{\infty} \, \sum_{m\, = \, 0}^{\infty} \, 
{2\,n+2\,m\choose n+m}  \, {n+m\choose n}^{2} \, 
{2\,n\choose n} \, {2\,m\choose m}
 \cdot \, x^n \, y^m \, \, \, \, \, = 
 \nonumber \\
\hspace{-0.9in}&& 
 \, \,\,  \, = \, \, \, \sum_{m\, = \, 0}^{\infty} \,  {2\,m\choose m}^2 \cdot \,
 _3F_2\Bigl([{{1} \over {2}}, {{1} \over {2}}\, +m, {{1} \over {2}}\, +m], \,
 [1, \, 1]; \, 16 \, y\Bigr) \cdot \, x^m
 \nonumber \\
\hspace{-0.9in}&& 
  \,\,\,  \,  = \, \, \, 
1 \,\, + 4  \,\, (x+\, y)\,\, +(36\, (x^2\, +y^2) +96\, x\, y)\, 
+[2160\,\, (x^2\, y  \, + x\, y^2) \, +400\,\, (x^3+y^3)]\, 
\nonumber \\
\hspace{-0.9in}&& \qquad \quad \,\,
+[4900\,\, (x^4+\, y^4)\, +44800\,\, (x\, y^3\,+\, x^3\, y)\, +90720\,\, x^2\, y^2]
 \,\,\,\, + \,\, \,\cdots 
\end{eqnarray}
This hypergeometric double series is solution of the
Picard-Fuchs system of PDEs
\begin{eqnarray}
\label{PDE13}
\hspace{-0.3in}&& \Omega_x \,\, = \,\, \, \, 
\theta^3_x \, \,\,  - 4  \, x \cdot \, (2\theta_x + 1)\, 
(\theta_x + \theta_y + 1)\, (2\theta_x + 2\theta_y + 1),
\nonumber \\
\hspace{-0.3in}&& \Omega_y\, \, = \,\,\,\,  
\theta^3_y  \,\,\,  - 4 \,  y \cdot \, (2\theta_y + 1)\, 
(\theta_x + \theta_y + 1) \, (2\theta_x + 2\theta_y + 1).
\end{eqnarray}

In the $\, y \, = \, x$ limit, this series reduces to the series 
\begin{eqnarray}
\label{190120}
\hspace{-0.9in}&&\sum_{n=0}^\infty \, \Bigl[ {2n \choose n} \, 
\sum_{k=0}^n {n \choose k}^2 {2k \choose k}{2n-2k \choose n-k}\Bigr] \cdot \, x^n 
 \,\, \,\,  = 
 \nonumber \\
\hspace{-0.9in}&& \qquad 
 \,\, =\,\,\,\,\, \,  \sum_{n=0}^\infty \,  {2n \choose n}^2  \cdot 
\,  _4F_3\Bigl([{{1 } \over {2}}, \, -n,\,  -n,\,  -n],
\,  \,[1,\,  1,\,  -\, {{2\, n \, -1 } \over {2}}]; \, 1\Bigr) \cdot \, x^n 
\nonumber \\ 
\hspace{-0.9in}&& \qquad \qquad 
 \,\, =\,\,\,\,\, \,  
 1 \, \,  \,+8\,x\,\,  +168\,{x}^{2}\,\,  +5120\,{x}^{3}\,
 +190120\,{x}^{4}\,\,  +7939008\,\,{x}^{5}\,\,  
 \\
\hspace{-0.9in}&&\qquad \qquad \quad \quad  \quad  \, \, \,  
+357713664\,{x}^{6}  +16993726464 \, \,{x}^{7}\,  \, \,
 + 839358285480\,\, x^8 \,\, + \,\, \,\cdots \nonumber
\end{eqnarray}
annihilated by the order-four Calabi-Yau operator:
\begin{eqnarray}
\label{defBatyrev2}
\hspace{-0.4in}&&\theta^4 \,  \,  \, \,-4 \, x \cdot 
(5\, \theta^2 \, +5 \,\theta \, +2) \cdot   (2\, \theta \, +\, 1)^2 \,
\,  \nonumber \\
\hspace{-0.4in}&&\quad \quad  \quad  \qquad  \quad +64 \, x^2 \cdot  
   (2\, \theta \, +\, 3) \cdot   (2\, \theta \, +\,1) \cdot
   (2\, \theta \, +\, 2)^2.  
\end{eqnarray}

The recast of the PDE system for the double series (\ref{24}) 
into the form (\ref{form}), gives two $(x, y)$-symmetric 
linear differential operators of order {\em nine}.
The singularities of the two order-nine 
linear differential operators are respectively 
$\,\,\, x \, \cdot \, (1 \, -16\,x)  \, = \, 0\,\,$ 
and $\,\,\, y \, \cdot \, (1 \, -16\,y)  \, = \, 0\,\,$  
together with the quadratic condition:
\begin{eqnarray}
\label{S2xy}
\hspace{-0.5in}{\cal S}_2(x, \, y) \,\, = \, \, \,\,\, \,
 2^8 \cdot \,(x-y)^2\,\,\, -2^5 \cdot \, (y+x)\,\,\,\, +1 
\,\,\,\,\, = \,\,\,\, \, \, 0,  
\end{eqnarray}
which has the simple rational parametrization
\begin{eqnarray}
\hspace{-0.1in}\quad (x, \, y) \,\,\, = \, \, \,\,
\Bigl(\Bigl({{1} \over {8}} \, - \, u   \Bigr)^2, \, \, \,
 \Bigl( {{1} \over {8}} \, + \, u  \Bigr)^2 \Bigr).
\nonumber 
\end{eqnarray}
The singularities $\,{\cal S}_2(x, \,y) \,= \, 0$ 
are, here also, logarithmic, the local exponents being 
$\,0,\, 1,\, 1,\, 2, \,3, \,\cdots,\, 7$.

These two $(x,\, y)$-symmetric order-nine operators 
also factorize in exactly, the same way as (\ref{factoL9L6}), 
in three order-one operators and an order-six operator
 like (\ref{L6ordersix}). The exterior square of this order-six operator
is also of order fourteen (instead of the order fifteen one expects for
a generic irreducible order-six operator), and, again,
 this order-six operator is homomorphic to its adjoint with a relation 
similar to (\ref{Homoadjoint}),  
the head coefficient in the order-two intertwinner 
being replaced by 
$\, 2^8 \, (16\,x -16\,y \,+3)/{\cal S}_2(x, \, y) /(16\,x-1)/x^2$.
We also have relations similar 
to (\ref{wronski}) for the various wronskians.

\subsection{Another Picard-Fuchs system above
 the Calabi-Yau operator (\ref{defBatyrev2})}
\label{picardmorebat2}
\vskip .1cm

Note that the Picard-Fuchs system of two variables 
``above'' the Calabi-Yau operator (\ref{defBatyrev2})
  is {\em not unique}.
Other  $(x, \, y)$-symmetric series 
 reduce to the series (\ref{190120}) annihilated by (\ref{defBatyrev2}), 
for instance, 
the double series expansion:
\begin{eqnarray}
\label{poch}
\hspace{-0.9in}&&\sum_{n\, = \,  0}^{\infty} \, \sum_{m\, = \, 0}^{\infty}  
\,\, \,{64}^{n+m} \cdot \, \, 
{\frac { (1/2)_n^{3} \cdot \,
  (1/2)_m^{3} \cdot \, (1/2)_{m+n} }{ (1)_{n+m}^{3} \cdot \, n! \,\, \,m!}}
  \cdot \, x^n \, y^m  
 \\
\label{poch1}
\hspace{-0.9in}&& 
\, \, \,\,  \, \, = \, \, \, \, \,\sum_{m\, = \, 0}^{\infty} \, 
  \Bigl(
{{ ({{1} \over {2}})_m } \over {m!}}
\Bigr)^{4} \, \times  \,   \\
\hspace{-0.9in}&& 
\qquad \quad \quad  \, \, \, \, \, 
 _4F_3\Bigl([{{1} \over {2}}, \, {{1} \over {2}}, 
\, {{1} \over {2}}, \, {{1} \over {2}}\, +\, m], 
\, [m \, + \, 1, m \, + \, 1, m \, + \, 1]; \, 64 \, x\Bigr) 
\cdot \, (64 \,y)^m 
\nonumber 
\end{eqnarray}
\begin{eqnarray}
\hspace{-0.9in}&& \, \, \,  \, \, = \, \, \, \, \, 
1\, \, \,\,  +4 \cdot (y+x) \, \,\, 
 \, +3 \cdot [27 \cdot (x^2+y^2)  +2 \cdot x \cdot y]
\nonumber \\
\hspace{-0.9in}&&\quad \quad \quad   \quad \quad  +20 \cdot (y+x) \cdot
  [125 \cdot (x^2+y^2) \, -122 \cdot x \cdot y] \,  \,
\nonumber \\
\hspace{-0.9in}&& \quad \quad  \quad  \quad \quad   +35/16 \cdot [42875 \cdot (x^4+y^4)
+162 \cdot x^2 \cdot y^2
+500 \cdot x \, y \cdot (x^2 \, + \, y^2)] 
\nonumber \\
\hspace{-0.9in}&&\quad \quad  \quad  \quad  \quad
+63/4 \cdot (y+x) \cdot [250047 \cdot (x^4+y^4) \,\,
 -248332 \cdot x \, y \cdot (x^2+y^2)
\nonumber \\
\hspace{-0.9in}&&\quad \qquad    \qquad  \qquad  \qquad   \quad 
+248602 \cdot x^2 \, y^2] \, \,\,\,\,   + \, \cdots 
\nonumber 
\end{eqnarray}
where $(a)_n$ is the usual Pochhammer symbol. This series can be found 
in Guttmann and Glasser~\cite{Glasser} as a lattice Green function.
It can also be seen as the expansion of a   
{\em Kamp\'e de F\'eriet function}~\cite{Kamp,Appell,Gaveau,Gaveau2} 
 (see  \ref{Kampe}):
\begin{eqnarray}
\label{KampKamp}
\hspace{-0.4in}F^{(1,3,3)}_{(3,0,0)} \left( [ {{1} \over {2}}],\, 
 [{{1} \over {2}}, \, {{1} \over {2}}, \, {{1} \over {2}}],\,  
[{{1} \over {2}}, \, {{1} \over {2}}, \, {{1} \over {2}}];\,  [1, 1, 1],-,-; 
\,\, 64 \, x,\, \,  64 \, y \right).
\end{eqnarray}

The double series (\ref{poch}) is not a series with integer coefficients
 but it can be recast\footnote[2]{Such series are called {\em globally 
bounded}~\cite{Christol}.} into a series with {\em integer} coefficients
if one performs the simple rescaling 
$(x, \, y) \, \rightarrow \, (4\, x, \, \, 4\, y)$. One obtains:
\begin{eqnarray}
\hspace{-0.7in}&& 1\,\,\,\,   + 16 \cdot \, (x\, + y)\,\,\, 
 +[1296 \cdot (x^2+y^2) +96 \cdot x \cdot y] 
\nonumber \\
\hspace{-0.7in}&& \quad \,  +1280 \cdot (y+x) \cdot 
[125 \cdot (x^2+y^2) \, -122 \cdot x \cdot y]
 \\
\hspace{-0.7in}&& \quad  \quad \,
+[24010000 \cdot (x^4+y^4) \, +280000 \cdot x \, y \cdot (x^2+y^2) \,
 +90720 \cdot x^2 \cdot y^2]
\nonumber \\
\hspace{-0.7in}&&\quad  \quad   \, 
+16128 \cdot (y+x) \cdot 
[250047 \cdot (x^4+y^4) \, -248332 \cdot x \, y \cdot (x^2+y^2) \, 
\nonumber \\
\hspace{-0.7in}&& \quad \quad \qquad   \qquad  +248602 \cdot x^2 \cdot y^2] 
 \, \, \,\,\,\,  \, + \,\,\, \cdots \nonumber 
\end{eqnarray}

The recast of the PDE system for the double series (\ref{poch}) 
into the form (\ref{form}) gives two $(x, y)$-symmetric 
linear differential operators, now, of order {\em thirteen}.

The  singular varieties of the two 
order thirteen operators $\, {\tilde \Omega}_x$ and $\, {\tilde \Omega}_y$
are respectively\footnote[5]{ Note that the limit $\, y \, = \, x$ 
 of the Picard-Fuchs systems associated with 
(\ref{poch}), is {\em actually a singular limit}.} 
$\,\,\, x \, \cdot \, (x-y) \cdot \, (1 \, -64\,x) \, = \, 0\,\,$ 
and $\,\,\, y \, \cdot \, (x-y) \cdot \, (1 \, -64\,y) \, = \, 0$,  
together with a $(x,\,y)$-symmetric {\em genus-zero} biquadratic which reads:
\begin{eqnarray}
\label{tildeS2xy}
\hspace{-0.5in} {\tilde S}_2(x, \, y) \,\, = \, \, \,\,\,
2^{12}\cdot \, x^2\, y^2 \,\,\,
 -2^7 \cdot \,x\, y\cdot \,(y+x)\, \,\, +(x-y)^2  
 \,\,\,\,\, = \,\,\,\, \,\, \, 0. 
\end{eqnarray}
The local exponents at the singularities of the order thirteen partial
linear differential operators are {\em independent}
of $\, y$ (respectively $\,x$). 

This genus zero curve (\ref{tildeS2xy}) has the rational parametrization
(well-suited for series expansions near $(x, \, y)\, = \, \, (0, \, 0)$)
\begin{eqnarray}
\label{singpar}
\hspace{-0.3in}x(t) \,\, = \,\,\,\, u^2, 
\qquad \quad  \,\,\,\, 
y(t) \,\, = \,\,\,\, \Bigl({{ u} \over { 1 \, + \, 8 \, u}} \Bigr)^2, 
\end{eqnarray}
or the  rational parametrization
\begin{eqnarray}
\hspace{-0.3in}x(u) \,\, = \,\,\,\, \Bigl({{u\, + \, 1} \over {8}}\Bigr)^2 , 
\qquad  \,\,\,\, 
y(u) \,\, = \,\,\,\, \Bigl({{u\, + \, 1} \over {8 \, u}}\Bigr)^2
 \,\, = \,\,\,\, x\Bigl({{1} \over {u}}\Bigr), 
\end{eqnarray}
the Atkin-Lehner-like involution 
$\, \, u \, \, \leftrightarrow  \,\, 1/u \, \, $ 
suggesting a modular curve interpretation of (\ref{tildeS2xy}).

Note that the two singular varieties 
$\, {\tilde S}_2(x, \, y)$ and $\, {\cal S}_2(x, \, y)$
(see (\ref{S2xy})),  are related by a simple involution:
\begin{eqnarray}
\label{tildeS2xyS2}
\hspace{-0.2in}{\tilde S}_2(x, \, y)\,   \,\, = \,\,\,\, \, \,\, 
2^{12} \cdot\,  \,x^2 \, y^2 \cdot \, \,  
{\cal S}_2\Bigl({{1} \over {2^{10} \, x}}, \, {{1} \over {2^{10} \, y}} \Bigr).
\end{eqnarray}
 
We thus see that the various Picard-Fuchs systems ``above'' a given Calabi-Yau ODE,
(i.e. reducing, when one takes the ``diagonal'' $\, y\, = \, x$,
to the same Calabi-Yau ODE), {\em do not have necessarily the same singular
manifolds}, even if these various singular manifolds must reduce to the
same singular points in the  $\, y\, = \, x$ limit. Since the
 singular variety (\ref{tildeS2xy}) contains the origin 
$(x,\,y) \, = \, \, (0, \, 0)$, it is easy to find, using  the  parametrization
(\ref{singpar}), a linear differential ODE satisfied by (\ref{poch}) 
when restricted\footnote[1]{See also the notion of Fuchsian 
system of linear partial differential
equations {\em along a submanifold} (see~\cite{Oaku}, in 
particular paragraph 6).} to the singular 
variety (\ref{tildeS2xy})
 (see (\ref{calC4}) in \ref{Kampe}). This cannot be done 
for (\ref{S2xy}) which does not contain the origin 
$(x,\,y) \, = \, \, (0, \, 0)$.

\vskip .1cm 

Breaking the $\, (x, \, y)$-symmetry in (\ref{poch}), by resumming  
the series as (\ref{poch1}), corresponds to the 
viewpoint of seeing Kamp\'e-de-Feriet
functions of several complex variables as
 {\em straight generalization\footnote[3]{The parameters of the 
hypergeometric functions become linear differential operators~\cite{Gaveau,Gaveau2}.}
 of hypergeometric functions}~\cite{Kamp,Appell,Gaveau,Gaveau2}. The 
$\, x$-singularities in each of the (transcendental) $\, _4F_3$ coefficients 
of the $\, y$-expansion (\ref{poch1}) are only the well-known 
$\, x\, =\,0$, $\, x\, =\,1$,  $\, x\, =\, \infty$ singularities of 
hypergeometric functions (here $\, x\, =\,1$ becomes
  $\, x\, =\,1/64$), and  are, of course, drastically different 
from the singular variety (\ref{tildeS2xy})
 for the double series (\ref{poch}).

\vskip .1cm 

The results for (\ref{poch}), can be generalized to more general 
(Kamp\'e de F\'eriet) double series depending on several parameters.
\begin{eqnarray}
\label{Kampdeftext}
\hspace{-0.3in}K(x, \, y) 
\, \,  \, = \, \, \, \,  \,  \,  \, 
\sum_{n\, = \, 0}^{\infty} \, \sum_{m\, = \, 0}^{\infty}  
\, {{(\alpha)_n^M \cdot (\beta)_m^M \cdot (\beta')_{m+n} } \over {
(\gamma)_{m+n}^M \, \, n! \, \, m! }} \cdot x^n \cdot y^m, 
\end{eqnarray}
where $\, (\alpha)_n$ is the usual Pochhammer symbol.
The same calculations as before show that 
their singular
curves {\em do not depend on the parameters}. 
These calculations for (\ref{Kampdeftext})
are displayed in \ref{Kampe}. 
 
\vskip .1cm 

\subsection{Picard-Fuchs systems with more than two 
variables ``above'' the Calabi-Yau operator (\ref{defBatyrev2}).}
\label{morevariables}

For heuristic reasons, we restricted to two variables
but one can find many Picard-Fuchs systems, with more than two complex
variables, ``above'' a given Calabi-Yau 
ODE like (\ref{defBatyrev2}). For instance,
the series (\ref{190120}) of the Calabi-Yau operator (\ref{defBatyrev2}) 
can also be written as the $\, x \, = \, y \, = \, z \, = \, t$  subcase
 of the (hypergeometric) series of four complex variables~\cite{Batyrev}:
\begin{eqnarray}
\hspace{-0.6in}
 \sum_{j, k, l, m} \,  \,\Bigl[ {2(j+k+l+m) \choose j+k+l+m} \cdot \,
\, \Bigl( {\frac{(j+k+l+m)!}{j! \, k! \,  l! \,  m!}} \Bigr)^2 \Bigr] 
\, \cdot     \,x^j \, y^k \, z^l \, t^m. 
\end{eqnarray}
The general term being hypergeometric, one obtains directly a system of four PDEs,
from which we build a linear ODE in the variable $x$,
 with $y$, $z$ and $t$ as "parameters".
Once one has series with four variables, and systems of PDEs with four variables,
one can take many limits in order to reduce to two variables.

For instance, if one restricts the previous series to $\, y \, = \, z \, = \, t$, 
one gets a series of two variables (which 
will of course reduce, for $\, y \, = \, x$, to the 
series (\ref{190120}) of the Calabi-Yau 
operator (\ref{defBatyrev2})), but is no longer symmetric in $\, x$ and $\, y$.
The series can be written as:
\begin{eqnarray}
\hspace{-0.7in}\sum_{N=0}^\infty \, {2N \choose N} \cdot \, 
 _3F_2 ([-N,-N,1/2],[1,1];4) \cdot \, 
 _2F_1 ([N+1, \, N+1/2], \, [1]; \, 4 \, x)
\cdot \, y^N  
\nonumber 
\end{eqnarray}
\begin{eqnarray}
\hspace{-0.7in}&& \qquad \, = \, \, \, \, \, \,
\sum_{N=0}^\infty\, {2N \choose N}^2 \cdot \, 
 _3F_2 ([-N,-N,-N],[1,1/2-N];1/4) \, 
 \nonumber \\
\hspace{-0.7in}&&  \qquad  \qquad \qquad  \times  \,  
_2F_1 ([N+1, \, N+1/2], \, [1]; \, 4 \, x)
\cdot \, y^N  
\nonumber 
\end{eqnarray}
\begin{eqnarray}
\hspace{-0.7in} \qquad\, = \, \, \, \, \, \,
 \sum_{n=0}^\infty\, \sum_{m=0}^\infty\, {{(2 n + 2 m)!}\over{(n! \, \,  m!)^2}} 
 \; \cdot \, _3F_2  ([-m,-m, 1/2],\, [1, 1],\,  4)
\, \cdot x^n \, y^m. 
\end{eqnarray}

The corresponding system of PDEs reads:
\begin{eqnarray}
\label{asym}
\hspace{-0.6in}&&\Omega_x \,\,\, = \,\,\, \,\, \, \theta^2_x\,\, \, \,
-\, 2\,x \cdot  \,(\theta_x + \theta_y + 1)\,(2\theta_x + 2\theta_y + 1),
\nonumber \\
\hspace{-0.6in}&&\Omega_y \,\,\, = \,\,\,\,\,\,  \theta^4_y\,\, \,\,
-\, 2\,y \cdot \, (10\theta^2_y + 10\theta_y + 3)\,
(\theta_x + \theta_y + 1)(2\theta_x + 2\theta_y + 1)\,
 \\
\hspace{-0.6in}&& \qquad  \quad 
 +\, 36\,y^2 \cdot \,(2\theta_x + 2\theta_y + 3)\,
(2\theta_x + 2\theta_y + 1)\, (\theta_x + \theta_y + 2)(\theta_x + \theta_y + 1).
\nonumber
\end{eqnarray}

Again, one can recast this system into a form like (\ref{form}), i.e. 
two linear differential operators $\, {\tilde \Omega}_x$
and $\, {\tilde \Omega}_y$ in the variable $\, x$ (resp. $\, y$), 
both of {\em order eight}, each one with the same singular variety 
which is the union of the two genus-zero 
algebraic curves:
\begin{eqnarray}
\hspace{-0.4in}&&16 \, x^2 \,\, - \, 8 \cdot \, (4 \, y \, +1) \cdot \, x \,\,
 + \, (4 \, y \, -1)^2 \, \,\,\, = \, \,\, \, 0, \qquad \qquad \hbox{and:}
\nonumber \\
\hspace{-0.4in}&&16 \, x^2 \,\, - \, 8 \cdot \, (36 \, y \, +1)\cdot  \, x \,\, 
+ \, (36 \, y \, -1)^2 \, \,\,\, = \, \,\, \, 0.
\end{eqnarray}
These two order-eight operators both factorise in a similar way as (\ref{factoL9L6}) 
but, this time, in the product of {\em two} order-one 
and one order-six operator. These two order-six operators rightdividing respectively
 $\, {\tilde \Omega}_x$ and $\, {\tilde \Omega}_y$ are not related 
by a $\, (x, \, y)$-symmetry, because the Picard-Fuchs system (\ref{asym}) 
is not $\, (x, \, y)$-symmetric.
Again these two order-six operators are such that their exterior square
are of order fourteen instead of the order fifteen one can expect 
for the  exterior square of a generic irreducible order-six operator.
Furthermore one has, again, that these order-six operators
are homomorphic to their adjoint, the intertwinner being of order two
(see (\ref{Homoadjoint})).  

\vskip .1cm 

More examples of Picard-Fuchs system with two variables ``above'' Calabi-Yau ODEs 
are sketched in \ref{morepic}, their corresponding (simple) singular varieties
being also given. 

\vskip .1cm 

\section{Singular manifolds for hypergeometric series 
of several complex variables}
\label{Horn}

All these singular varieties (\ref{cand}), (\ref{singsing}), (\ref{tildeS2xy}) 
(as well as similar ones, (\ref{singBat5}), (\ref{singBat6}), given 
in  \ref{morepic}) can, in fact, be easily obtained from very 
simple calculations when one remarks that the previous double series are 
{\em hypergeometric series of several complex 
variables}. The calculations, corresponding to the Horn's convergence theorem,
 are similar to the ones for Horn functions 
and Horn systems~\cite{Horn,Horn2,Horn3,Horn6}. A very important property 
is the fact that the region of convergence for hypergeometric series {\em does 
not depend on the parameters}~\cite{Srivastava}. 

Let us denote  the coefficients of (\ref{H2}), by $\, c_{n, \, m}$
\begin{eqnarray}
 c_{n, \, m} \, \, \, = \, \, \, \, {{ (3m+3n)! } \over { n!^3 \, \, m!^3 }}, 
\end{eqnarray}
the successive ratio of $\,  c_{n, \, m}$ in the two ``directions'' 
reads respectively 
\begin{eqnarray}
\label{ratioCnm}
\hspace{-0.4in}&&{{c_{n, \, m}} \over { c_{n+1, \, m}}}
 \,\, = \,\,\,
 {{ (n+1)^3} \over { b(n, \, m) }},
 \qquad \qquad {{c_{n, \, m}} \over {c_{n, \, m+1}}} \,\, = \,\,\, 
{{ (m+1)^3} \over { b(n, \, m) }}, 
\end{eqnarray}
where the product $\, b(n, \, m)$ is given by (\ref{defbnm}).
In the $\, n$ and $\, m$ large limits 
these two ratios behave respectively like 
\begin{eqnarray}
\label{39}
\hspace{-0.7in} X(n, \, m) \,\, = \, \,\,
 {{n^3} \over{27 \, (m\, +n)^3}} ,
 \qquad \hbox{and} \qquad  \, \,
 Y(n, \, m) \,\, = \, \,\, {{m^3} \over{27 \, (m\, +n)^3}},   
\end{eqnarray}
where one remarks that $\, X(n, \, m)$ and  $\, Y(n, \, m)$ 
depend only of the ratio $\, n/m$. 
The curve rationally parametrized by 
$\, (x, \, y) \, = \, \, (X(n, \, m), \, Y(n, \, m))$
can easily be obtained performing a resultant (elimination
of $\, m$ or $\, n$ or the ratio $\, n/m$) and one recovers, 
in a very simple way the singular manifold (\ref{cand}).  One 
notes that (\ref{39}) is nothing but the previous
 binary cubics (\ref{cubic}) yielding (\ref{discrim}),
the discriminant of a two-parameters family of Calabi-Yau 3-folds.

We can perform similar calculations for the hypergeometric series (\ref{24}), 
the ratio of the  $\, c_{n, \, m}$'s 
also read (\ref{ratioCnm}), the product  $\, b(n, \, m)$ being now given by 
\begin{eqnarray}
\hspace{-0.4in}b(n, \, m)  \,\,\, = \,\,\,\,\,
2 \cdot \, (2\,n \, +2 \, m \, + 1) \,\, (2\,n \, +2 \, m \, + 2) \, \, (2\,n \, + 1).
\end{eqnarray}
In the $\, n$ and $\, m$ large limit, this gives the rational parametrization
of the singular variety (\ref{S2xy}),  namely 
$\, (x, \, y) \, = \, \, (X(n, \, m), \, Y(n, \, m))$,  with:
\begin{eqnarray}
\label{42}
\hspace{-0.7in} X(n, \, m) \,\, = \, \,\, {{n^2} \over{16 \, \,  (m\, +n)^2}} ,
 \qquad \hbox{and} \qquad  \quad 
 Y(n, \, m) \,\, = \, \,\, {{m^2} \over{16 \, \,  (m\, +n)^2}}.   
\end{eqnarray}

For the hypergeometric series (\ref{poch}), 
the ratio of the  $\,  c_{n, \, m}$'s  read respectively
\begin{eqnarray}
\label{ratioCnmbis}
\hspace{-0.6in}&& {{ (n+m+1)^3 \, (n+1)} \over {
 4  \cdot \, (2\, n\, +1)^3 \, (2\, n\,+ 2\, m\, +1)  }},
 \quad \quad \quad 
{{  (n+m+1)^3 \, (m+1)} \over {  4 \cdot \, (2\, m\, +1)^3 \, (2\, n\,+ 2\, m\, +1)}}. 
 \nonumber 
\end{eqnarray}
In the $\, n$ and $\, m$ large limits this gives 
 the rational parametrization
of the singular variety (\ref{tildeS2xy}),  namely 
$\, (x, \, y) \, = \, \, (X(n, \, m), \, Y(n, \, m))$,  with:
\begin{eqnarray}
\label{XYnm2}
\hspace{-0.5in} X(n, \, m) \,\, = \, \,\, {{(m\, +n)^2} \over{64 \, n^2}} ,
 \qquad \hbox{and} \qquad \quad 
 Y(n, \, m) \,\, = \, \,\,  {{(m\, +n)^2} \over{64 \, m^2}} .   
\end{eqnarray}
 
\vskip .1cm 
Finally for other hypergeometric series (\ref{defBat5}), 
(\ref{defBat6}), given in  \ref{morepic}, similar 
calculations also 
give  rational parametrizations of the corresponding {\em genus-zero} 
singular curves (\ref{singBat5}) and (\ref{singBat6}).

For instance the successive ratio of $\, c_{n,\, m}$'s 
for (\ref{defBat6}) read respectively 
\begin{eqnarray}
\hspace{-0.2in}{\frac { (n+1)^{4}}{ b(n, \, m)  }}, 
\qquad \quad 
{\frac { \left( m+1 \right)^{4}}{ b(n, \, m) }}, 
\qquad \qquad   \hbox{where:} 
\end{eqnarray}
\begin{eqnarray}
\hspace{-0.5in}b(n, \, m) \,\, = \, \,\, \, \,
(2\,n+m+1)  \, (2\,n \,+m+2)  \, (2\,m\, +n+1)  \, (n\,+m+1). 
\nonumber 
\end{eqnarray}
In the $\, n$ and $\, m$ large limit this gives 
 the rational parametrization
of the singular variety (\ref{singBat6}),  namely 
$\, (x, \, y) \, = \, \, (X(n, \, m), \, Y(n, \, m))$,  with:
\begin{eqnarray}
\label{XYnm3}
\hspace{-0.5in} X(n, \, m) \,\, = \, \,\, 
{\frac {{n}^{4}}{ (2\,n\,+m)^{2} \, (2\,m\,+n)  \, (n\,+m) }},
 \quad  \quad 
 Y(n, \, m) \,\, = \, \,\, X(m, \, n).
\nonumber 
\end{eqnarray}
Of course, all these calculations can be performed with 
series of  {\em any finite} number of complex variables.
These (simple) calculations are  only valid
 for  series of several complex variables,
such that the ratio of the various consecutive coefficients (see (\ref{ratioCnm}))
are rational expressions (typically {\em hypergeometric series}).

\vskip .1cm 

\section{Towards singular manifolds of Ising model D-finite system of PDEs}
\label{towards}

One thus sees, from the previous calculations, that one can actually define, and find
without ambiguity, the
singular manifolds of  D-finite systems of PDEs.  The singular manifolds
are {\em fixed}, and can (in principle) be obtained from (possibly tedious
but well-defined) calculations from the D-finite system of PDEs.
This is quite different from the case of generic  (non-holonomic) systems of  
PDEs where singularities depend on initial boundary conditions. 
With the previous calculations, one can see that the singular manifolds can even be obtained 
from very simple calculations in the (selected) case of {\em hypergeometric series}, the 
singular varieties with {\em rational parametrization} being underlined. 

\vskip .3cm 

For functions of several complex variables which are not known to be solutions
of D-finite systems of partial linear differential operators (or even partial non-linear 
differential operators  but with fixed critical points), the question of defining and 
finding the singular manifolds seems hopeless. 
There is, however, one category of functions of several complex variables 
that emerges quite naturally in physics, where some hope remains, thus partially
justifying,  the ``guessing'' approach often
performed in lattice statistical 
mechanics~\cite{challenge,challenge2,challenge3,WuZia,Meyer,partitions,zeroWu,transmi}. 
These functions of several complex variables are the ones which can be 
decomposed as infinite {\em sums of D-finite functions} (in a
 typical Feynman diagram approach).
The best example is the full susceptibility of the anisotropic square Ising model
which has such a decomposition~\cite{wu-mc-tr-ba-76}. Let us try to find 
the singularity manifolds of the anisotropic $\, \chi^{(n)}$'s, trying
in a second step, to understand the singularity manifolds of the {\em anisotropic} 
full susceptibility  $\, \chi$.

\subsection{Landau approach for the singular manifolds of 
the anisotropic $\, \chi^{(n)}$}
\label{anisochin}

Finding the Fuchsian (and in fact globally 
nilpotent~\cite{bo-bo-ha-ma-we-ze-09}) linear ODEs 
for the $\, n$-fold integrals $\, \chi^{(n)}$'s of the decomposition of the 
 full magnetic susceptibility of the 
square lattice Ising model is already a ``tour-de-force'' 
in the isotropic case~\cite{Khi3,Khi4,ze-bo-ha-ma-05c,Khi6,High}.
 
The anisotropic $\, \chi^{(2)}$, has a surprisingly nice
factorized form (see equation (3.22) in~\cite{ongp}). It
is the product of the isotropic $\, \chi^{(2)}$ and of a simple square-root 
algebraic function:
\begin{eqnarray}
\chi^{(2)}(k, \, r) \, \,  = \, \, \,  \, \, 
 {{\Bigl((1+k\, r) \cdot (k\,+\,  r)\Bigr)^{1/2}} \over {1+k}} \cdot 
\chi^{(2)}(k, \, 1),
\end{eqnarray}
where $\, k\, = \, s_1\, s_2\, $ is the modulus of elliptic 
functions in the parametrization of the model, where the ratio
$\, r \, = \, s_1/s_2$ is the anisotropy variable, with
$\, s_1 \, = \, \, \sinh 2K_1$, $\, s_2 \, = \, \, \sinh 2K_2$, 
(with notations $\, K_1 \, = \, E^v/k_BT$ and 
$\, K_2 \, = \, E^h/k_BT$, see (3.22) of~\cite{ongp}),
and where $\, \chi^{(2)}(k, \, 1)$
is the isotropic $\chi^{(2)}$:
\begin{eqnarray}
\hspace{-0.6in}&&\chi^{(2)}(k, \, 1) \, \, = \, \, \, \, \, 
  {{1} \over {3 \, \pi}} \cdot {{
 (1+k^2) \cdot E(k^2) - (1-k^2) \cdot K(k^2) } \over {(1-k) \, (1-k^2)}}
\nonumber \\
\hspace{-0.6in}&& \qquad\qquad  \, \, = \, \, \, \, \, 
{{k^2} \over {4 \, (1\, +k)^4 }} \cdot \,
 _2F_1\Bigl([{{3} \over {2}}, \,{{5} \over {2}}],
 \, [3]; \, {{4 \, k} \over {(1\, +\, k)^4}}\Bigr).
\end{eqnarray}

Beyond this surprisingly
simple $\, \chi^{(2)}$ case, obtaining a D-finite (Picard-Fuchs)
 system for $\, \chi^{(3)}$,
for the anisotropic square Ising model, would require too massive and 
extreme computer calculations. Furthermore, the simple ``Horn calculations''
 detailed in section (\ref{Horn}) 
require some {\em closed asymptotic formula} (or  
some asymptotic formula of exact linear recursions) for the coefficients 
of the double series of the 
anisotropic $\, \chi^{(n)}$, and would require some assumption that 
the $\, \chi^{(n)}$'s are hypergeometric series, or at least, that their 
singular part is dominated by hypergeometric series.

However, if one is only interested in the singularities of 
such D-finite $\, n$-fold integrals, the {\em  Landau singularity approach},
 we have already used in the isotropic case, to find~\cite{Landau2,SingNfold}
 these singularities, can again, be worked out. We are not going to recall 
the details of this approach, which 
correspond in the anisotropic case, to sometimes quite tedious (algebraic) 
calculations. The idea, which is specific
of $\, n$-fold integrals of some algebraic integrands, amounts 
to saying that the singularities should, in principle, be deduced only from
the algebraic integrands of these integrals from elementary algebraic 
calculations~\cite{Eden,Landau,Landau2,SingNfold,Experimental}.
 
We will display, in a following 
subsection (\ref{resultanisochin}) the results for the first $\, \chi^{(n)}$'s 
 after recalling in the next subsection a first set of fundamental singularities. 

\subsection{Nickelian singular manifolds for the anisotropic $\, \chi^{(n)}$'s
and zeroes of the partition function}
\label{nickellian}

In contrast to the form factors~\cite{Holonomy,Assis} $C^{(n)}(M, \, N)$, 
whose only singular points
are $\, k\, = \, \, 0$, $\, k\, = \, \, 1$ and $\, k\, = \, \,  \infty$, 
the $\, \chi^{(n)}(k)$'s have many further singularities. The first set 
of these singularities was found, by Nickel~\cite{nic1,nic2}, 
to be, for the isotropic case 
($\, K_1 \, = \, \,K_2 \, = \, \,K$), located at 
\begin{equation}
\label{location}
\cosh^{2}2K \,\,\,\, -\,\sinh 2K \cdot \, (\cos(2\pi j/n)\,+\cos(2\pi l/n))
\,\,\,\, =\,\,\,\, \, 0, 
\end{equation}
with ($[x]$ being the integer part of $x$): 
$\, 0\,\leq\, j,\,\,\, \,  l \,\leq \, [n/2], \, 
 j=l=\,0$  excluded (for $n$ even, $j+l=\,n/2$ is
also excluded).  
Equivalently (\ref{location}) 
reads:
\begin{eqnarray}
\hspace{-0.8in}&& \sinh 2K_{j,l}\,\, =\, \,\, s_{j,l}
 \,\,\, =\, \,\,\,\, 1/2 \cdot \, (\cos(2\pi j/n)+\cos(2\pi l/n))\,  
\nonumber \\
\hspace{-0.8in}&&\qquad \qquad  \quad  \quad   \,\, 
\pm i/2 \cdot \, [(4-(\cos(2\pi j/n) \, +\cos(2\pi l/n))^2]^{1/2}.
\end{eqnarray}
\vskip .1cm 
These Nickel's singularities  are clearly
 on the unit circle 
$\, |s| \, =\, 1$, or $\, |k| \, =\, 1$. 
Do note that this is no longer the case for the anisotropic 
model where Nickel's singularities for the anisotropic  
$\chi^{(n)}$'s become: 
\begin{eqnarray}
\label{location2}
\hspace{-0.5in}&&\cosh2K_1 \cdot \, \cosh2K_2 \,\, \\
\hspace{-0.5in}&&\qquad \quad \quad  -\,(\sinh 2K_1\cdot \, \cos(2\pi j/n)\,
+ \sinh 2K_2 \cdot \,  \cos(2\pi l/n))
\,\, \,\,=\,\,\, \,\, \, 0, \nonumber 
\end{eqnarray}
with $\, j$,  $\, l\, = \, \, 1, \, 2, \, \cdots, \, n$. These (complex) 
algebraic curves (\ref{location2}), 
in the two complex variables
$\, s_1 \, = \, \, \sinh 2K_1$, $\,  s_2 \, = \, \,\sinh 2K_2$, have to be
singular loci (as will be suggested in the following section)
 for the D-finite system
of PDEs satisfied by the anisotropic (holonomic) 
$\, \chi^{(n)}$'s.

One can rewrite these algebraic curves in 
$\, k \, = \, \, s_1\cdot \, s_2$ 
and $\, r \, = \, \, s_1/s_2$ as
\begin{eqnarray}
\label{genus1}
\hspace{-0.2in} \quad (r+k) \, \cdot \, (k\, r+1)\,\, \, \,
-k \, \cdot \, (r \, \, U\, \pm \, V)^2
 \, \,\, \, = \,\,\, \,\, \, \, 0, 
\end{eqnarray}
where $\, U \, = \, \, \cos(2\pi j/n)$ and $\, V \, = \,  \,\cos(2\pi l/n)$.
Do remark that {\em these algebraic curves depend on the anisotropy variable} 
$\, r \, = \, \, s_1/s_2$. We will underline this important fact
in subsection (\ref{comments}).
Remarkably these curves are 
{\em generically of genus-one}\footnote[1]{For 
$\, U \, = \, \, V$ (as well as  
$\, U \, = \, \, -V$, $\, U \, = \, \pm 1$, 
$\, V \, = \, \pm 1$) the curves 
are {\em genus-zero}. For instance, for $\, U \, = \, \, V$, they read
$(r\, \pm \, 1)^2\, k \cdot \, U^2
  \, \,\, -(r+k)\cdot \, (k\,r+1)\, \,\,  = \,\,\,  \, 0$.}, not only when 
$\, U \, = \, \, \cos(2\pi j/n)$ and $\, V \, = \,  \,\cos(2\pi l/n)$,
but for {\em any fixed value of} $\, U$ and $\, V$.
Their $\, j$-invariant~\cite{SingNfold,broglie} reads\footnote[2]{
This rational expression (\ref{jUV}) of $\, U$ and $\, V$ is nothing but 
relation (36) in~\cite{broglie}
with $\, J_x/J_z\, = \, U$, $\, J_y/J_z\, = \, V$.
This rational expression
remarkably factorizes for many Heegner numbers~\cite{Heegner,Lario}
(complex multiplication cases): 
$\, j \, = \, \, 12^3, \, 20^3, \, (-15)^3, \, 2\cdot \, 30^3, \, 66^3$ 
and selected quadratic values of $\, j$-invariant,
 like $\, j^2+191025\, j \, -495^3 \, = \, \, 0$
or $\, j^2 \, -1264000 \, j \, -880^3 \, = \, \, 0$. This 
(partially) explains the occurrence in (\ref{location2}) 
of several complex multiplication cases
 (for instance $\, U \, = \, \, \cos(2\pi 2/8)$, 
 $\, V \, = \, \, \cos(2\pi /8)$ which give $\, j \, = \, \,1728$).
}:
\begin{eqnarray}
\label{jUV}
\hspace{-0.2in}j\,\,\,  = \, \, \, \,
256 \cdot \,{\frac { ({U}^{4}+{V}^{4}-{V}^{2}{U}^{2}-{U}^{2}-{V}^{2}+1)^{3}}
{ ({V}^{2}-1)^{2} \,\, ({U}^{2}-1)^{2}\,
 \, ({U}^{2}-{V}^{2})^{2}}}.
\end{eqnarray}
We thus see that we do have a {\em two-parameters family
of elliptic curves}.

These elliptic (or rational)  curves (\ref{location2}) accumulate with 
increasing values of $\, n$, in the same way Nickel's singularities
(\ref{location}) accumulate on the unit circle 
$\, |s| \, =\, 1$, in a certain (real) submanifold $\, {\cal S}$  of the 
two complex variables $\, s_1$,
$\, s_2$ (four real variables). However, this ``singularity
manifold'' $\, {\cal S}$ is not a codimension-one (real) 
submanifold (like the unit circle 
$\, |s| \, =\, 1$ in the $\, s$-complex plane), but 
actually a codimension zero submanifold, 
as can also be seen on various analysis of complex temperature zeroes 
(see\footnote[5]{The first reference corresponds to the fact that zeroes 
can fill areas in the complex temperature plane. Some later papers contain results 
on the density of zeroes in the thermodynamic limit. } 
for instance~\cite{Steph1,Steph2,Steph3,Steph4,Steph5,Steph6,Wood1,Wood2}
and more recently~\cite{wang-2002,saarloos-kurtze-1984,lu-wu-2001,matveev}). Note 
that this  ``singularity
manifold'' becomes very ``slim'' near the (critical) algebraic curve
$\, k \, = \, \, s_1  \, s_2 \, = \, \, 1$
(see for instance the region near the real axis of
 figures 1, 2 and 3 in~\cite{wang-2002}). 

In the isotropic case, we actually
 obtained~\cite{Khi3,Khi4,Experimental,Khi6,High,Khi5exact} 
the  linear ODEs satisfied by the first $\, \chi^{(n)}$'s, 
for $\, n\, = \,  \, 3, \, 4, \, 5, \, 6 \,$
and, thus, of course, the corresponding ODE singularities. Furthermore,
 we also performed
a {\em Landau singularity} approach that enabled
 us to obtain, and describe, the singularities
for {\em all}~\cite{Landau2,SingNfold} the $\, \chi^{(n)}$'s. These exact 
results show, very clearly, that there are (non-Nickelian) singularities 
inside the unit circle and outside the unit circle (see Figure 1, 2, 3 and 4 
in~\cite{SingNfold}). On the figures of~\cite{SingNfold} it is easy to get
convinced that the accumulation of these non-Nickelian singularities  
will probably be a {\em dense set of points inside the unit circle}
and (by Kramers-Wannier duality) {\em outside the  unit circle}.
These non-Nickelian singularities are given in terms 
of Chebyshev polynomials of the first and second kind (see equations (28) and (29) 
in~\cite{SingNfold}). Upgrading these slightly involved exact (Chebyshev) 
non-Nickelian results~\cite{SingNfold} for the isotropic model
 to the anisotropic model is,
at the present moment, probably too ambitious. 

Let us simply try, using the 
previous {\em Landau singularity} approach, to provide, may be 
not an exhaustive description of all the singularities for the anisotropic case,
but at least, the exact expression of all the singular manifolds 
(Nickelian or non-Nickelian) for the first  anisotropic $\, \chi^{(n)}$'s.

\subsection{Singular manifolds for the first  anisotropic $\, \chi^{(n)}$}
\label{resultanisochin}

The {\em Landau singularity} approach detailed in~\cite{Landau2,SingNfold}
for the isotropic $\, \chi^{(n)}$'s of the square Ising model, can easily be 
generalized to the anisotropic $\, \chi^{(n)}$'s. We are not going to explain 
here the details of these (slightly tedious) calculations which are 
basically the same as in~\cite{Landau2,SingNfold} mutatis mutandis.
The calculations being slightly involved we just give the results for the 
first $\, \chi^{(n)}$'s. 

The singularities of $\, \chi^{(3)}$ and $\, \chi^{(4)}$ read
respectively in $\, k$ and $\, r$:
\begin{eqnarray}
\label{sing3}
\hspace{-0.8in}&& Sing(\chi^{(3)}) \,  \, \, = \, \, \,  \, \, 
(k^2-1) \cdot (3\,kr+r+4\,{k}^{2})  \cdot ({k}^{2}r+3\,kr+4)  \cdot
 ({k}^{2}r+r+k) \nonumber \\
\hspace{-0.8in}&&  \qquad  \quad  \quad \times  (3\,{r}^{2}k-r-k-{k}^{2}r)  \cdot 
( 4+3\,kr+4\,k+4\,{k}^{2})  \, (r+k)  \, (kr+1), \\
\label{sing4}
\hspace{-0.8in}&&Sing(\chi^{(4)}) \,  \, \,  = \, \,  \, \, \, 
 (k^2-1) \cdot  (kr+1+{k}^{2}) 
 \cdot (3\,{r}^{2}k-r-k-{k}^{2}r).
\end{eqnarray}
In order to compare these results with our previous exact results
for the isotropic model, which were given~\cite{Khi3,Experimental,Khi6} 
in the (quite natural for 
such $\, n$-fold integrals) variable~\cite{Khi4,Khi6} $\, w$, let us 
rewrite these results in $\, r$ and  $\, w\, = \, \, s/(1+s^2)/2$, 
where, now,  $\, s\, = \, \, (s_1 \,s_2)^{1/2} $: 
\begin{eqnarray}
\label{singw}
\hspace{-0.8in}&& Sing(\chi^{(3)}) \, \, = \, \, \,\,\, 
  (w^2-1)  \cdot {w}^{2} \cdot
 ( {r}^{2}-4\,r+4+3\,{w}^{2}{r}^{2}-4\,{w}^{2}r+16\,{w}^{4}r)^{2}  
\nonumber \\
\hspace{-0.8in}&&  \qquad \quad  \quad \times  
\, (1+4\,{w}^{2}r-2\,r)^{2} \, ( 3\,{r}^{2}-1-4\,{w}^{2}r+2\,r)^{2}
\nonumber \\
\hspace{-0.8in}&&  \qquad \quad \quad   \times  
  (3\,r-4+16\,{w}^{2})^{2} \cdot (1+4\,{w}^{2}r-2\,r+{r}^{2})^{2},   \\
\label{singwbis}
\hspace{-0.8in}&&Sing(\chi^{(4)}) \, \, \, = \, \, \, \, \,
{w}^{2} \cdot (w^2-1)  \cdot \,
 \, (4\,{w}^{2}-2+r)^{2} \cdot (3\,{r}^{2}-1-4\,{w}^{2}r+2\,r)^{2}.
\end{eqnarray}
Note that the complex multiplication points 
of the isotropic case~\cite{SingNfold},
namely the roots of $\, 1\, +\, 3\, k\, + \, 4 \, k^2 \, = \, \, 0$
and  $\, k^2\, +\, 3\, k\, + \, 4 \,  \, = \, \, 0$, 
come from the $\, Sing(\chi^{(3)})$ factor
\begin{eqnarray}
\label{factor1}
\hspace{-0.3in}{r}^{2}\, -4\,r\,+4\,+3\,{w}^{2}{r}^{2}\,-4\,{w}^{2}r\,+16\,{w}^{4}\, r, 
\end{eqnarray}
in (\ref{singw}),  or equivalently with $\, (k, \, r)$, the two factors
 in (\ref{sing3}):
\begin{eqnarray}
\label{factor2}
\hspace{-0.3in}(3\,kr\,+r\,+4\,{k}^{2})  \cdot \, ({k}^{2}r \, +3\,kr\,+4),
\end{eqnarray}
The vanishing condition of (\ref{factor1}) corresponds to a 
 {\em genus-zero curve}, its rational
parametrization  being:
\begin{eqnarray}
w \, = \, \, \, \,{\frac {{u}^{2}+1}{2 \, u}}, 
\qquad \,\,\,\,
r \, = \, \,  \,{\frac {-4}{ \, {u}^{2} \cdot ({u}^{2}+3) }}.
\end{eqnarray}

Note that  $\, \, Sing(\chi^{(3)})$ and  $\, Sing(\chi^{(4)})\, $ 
have a non-trivial gcd (respectively in $\, k$, then $\, w$):
\begin{eqnarray}
\hspace{-0.9in}&&
gcd(Sing(\chi^{(3)}), \,Sing(\chi^{(4)})) 
\,\,  = \, \, \,\, 
  (k^2-1) \cdot \,  (3\,{r}^{2}k-r-k-{k}^{2}r),
\nonumber \\
\hspace{-0.9in}&&
gcd(Sing(\chi^{(3)}), \,Sing(\chi^{(4)}))
 \,\,  = \, \, \,\, 
 {w}^{2} \cdot (1-w) 
 \, (1+w)  \cdot (3\,{r}^{2}-1-4\,{w}^{2}r+2\,r)^{2}, 
\nonumber
\end{eqnarray}
the last algebraic curve 
$\, 3\,{r}^{2}k\,-r-k \,-{k}^{2}r \, = \, \, 0$,
is a  {\em genus-one curve}. 
A way to understand, in the anisotropic case,  the emergence 
of singular algebraic curves shared by several $\, \chi^{(n)}$'s 
($\, n$ even and $\, n$ odd) amounts to noticing that 
these curves actually reduce, in the isotropic limit, to 
$\, k\, = \, \, 1$, the singular variety of the partition function 
of the anisotropic model.

\vskip .1cm 

The fact that the singular curve $\, 3\,{r}^{2}k-r-k-{k}^{2}r \, = \, \, 0$,
together with the Nickelian
algebraic curves (\ref{location2}), (\ref{genus1}),  
are {\em not genus-zero} (as all the genus-zero curves 
of section (\ref{picardmore}), like (\ref{S2xy}), (\ref{tildeS2xy}), 
as well as the ones displayed in  \ref{morepic}, 
see (\ref{singBat5}), (\ref{singBat6})),
show that the series for the anisotropic $\, \chi^{(n)}$'s 
 {\em cannot be hypergeometric series} in the variables $\, k$ and $\, r$ 
(see section (\ref{Horn})). 
\vskip .1cm 
It would be interesting, before trying to generalize the Chebyshev polynomial
formula~\cite{SingNfold} for the non-Nickelian 
singularities of the isotropic model, 
to the anisotropic one, to accumulate, with this Landau singularity 
approach, more non-Nickelian algebraic curves in the anisotropic case.
Recalling the systematic emergence of elliptic curves (see (\ref{genus1}))
for the  Nickelian algebraic curves, it would be interesting 
to {\em systematically look at the genus of these singular curves}, to see if higher 
genus curves are also discarded for the non-Nickelian algebraic curves. 

\vskip .1cm 
It would be also interesting to confirm these Landau singularity 
calculations, with differential algebra calculations. Even with the last progress
performed by Koutschan on the {\em creative telescopic}
method~\cite{Koutschan,Koutschan2}, getting the (Picard-Fuchs) 
system of PDEs satisfied by the several complex variables
series of the anisotropic  $\, \chi^{(n)}$'s corresponds, at the present moment,  
to too large calculations (even for the anisotropic $\, \chi^{(3)}$).
However, if one considers particular anisotropic subcases 
($\, s_2\, = \, \, 3 \, s_1$, $\, s_2\, = \, \, 5 \, s_1^2$, ...),
obtaining the corresponding ODEs for the anisotropic $\, \chi^{(3)}$,  
in the unique complex variable, could be imagined using the 
creative telescopic method~\cite{Koutschan,Koutschan2}, or even,  
from series expansion as we did in the isotropic case~\cite{Khi3}. 

\subsection{Singular manifolds and the anisotropy variable}
\label{comments}

For experts of Yang-Baxter integrability, the fact that the singularities varieties, 
namely the Nickelian elliptic curves (\ref{genus1}), or the non-Nickelian rational 
curves (\ref{factor2}), {\em do depend on the anisotropy} of the model may come as 
a surprise. Indeed, within the Yang-Baxter integrable framework, and as a consequence 
of the existence of families of commuting transfer matrices (row-to-row, diagonal or 
corner transfer matrices), one  used to have many quantities like the order parameter, 
the eigenvectors of row-to-row or corner transfer matrices, ..., which are {\em independent}
 of the so-called ``spectral parameter'' (the parameter that enables to move along each 
elliptic curve). The selected quantities {\em depend only on the modulus} $\, k$ of 
the elliptic functions. Along this line, one certainly expects the singular manifolds, 
which are highly symmetric, ``invariant'' and ``universal'' 
manifolds~\cite{sixteen,inversionWu,inversionRollet}, to be 
{\em also independent of the spectral variables}.  With 
the previous variables $\, k$ and $\, r$, the singular manifolds 
should just depend on the modulus $\, k$, and {\em not on the anisotropy variable} 
$\, r$ (related to the spectral parameter). 
The surprise is that the singular manifolds do depend also on the 
anisotropy variable $\, r$, and thus on the spectral variable.

The $\, \chi^{(n)}$'s are known~\cite{Holonomy} to be an infinite sum
of  form factors $\, C^{(n)}(N,\, M)$:
\begin{eqnarray}
\label{finite}
\chi^{(n)} \,\,  \, = \, \, \, \, \, \, \, \sum_M \, \sum_N \,C^{(n)}(N,\, M),  
\end{eqnarray}
this relation being inherited from the fact that the full susceptibility 
is the sum of all the two-point correlation functions~\cite{Holonomy}.

Recalling the simplest (nearest neighbour) correlation 
function $\, C(0,1)$, it reads~\cite{FuchsPainleve} 
in the anisotropic case\footnote[5]{We use the 
maple notations for $\, \Pi$ and $\, K$.}:
\begin{eqnarray}
\label{lets}
\hspace{-0.7in}&& C(0,1) \,\,   \,   = \, \, \, \, \,  \, \, 
{{2} \over {\pi \, r}} \cdot \, 
\Bigl({{k+r} \over {k}}\Bigr)^{1/2} \cdot \,  
\Bigl(
(1\, + \, k\, r) \cdot \, \Pi(- \, k \, r,\, k) \, - \, \, K(k)
\Bigr), 
\nonumber
\end{eqnarray}
where, again, $\, k\, = \, s_1 \, s_2$ is the modulus of the 
elliptic functions parametrizing
the model, and $\, r$ is the ratio $\, r \, = \, s_1/s_2$ 
and where $\,\Pi(x, \, y)$ 
is the complete elliptic integral
of the third kind. 

The singular manifolds correspond to the singular points 
of the complete elliptic integrals 
of the first and third kind, namely $\, k\, =\,  0$, $\, k\, = \, 1$ 
and $\, k\, = \, \infty$. Therefore they depend only on the modulus $\, k$ in 
the elliptic parametrization of the model. 

The form factors have been seen to be solutions of linear differential equations
associated with elliptic functions~\cite{Holonomy,Assis}. Consequently, their 
singular points correspond to the singular points of the complete elliptic integrals 
of the first or second kind $\, E$ or $\, K$, namely $\, k\, =\,  0$, $\, k\, = \, 1$ 
and $\, k\, = \, \infty$. The generalization to the anisotropic case has been sketched
in~\cite{FuchsPainleve}. One expects the results to be polynomial 
expressions of  the complete elliptic integrals 
of the first (or second) and third kind, yielding again, singular manifolds 
which depend only on the modulus $\, k$, and are actually
 $\, k\, =\,  0$, $\, k\, = \, 1$, 
or $\, k\, = \, \infty$.

Finite sums of correlation functions or form factors, 
certainly have  $\, k\, =\,  0$, $\, k\, = \, 1$ 
or $\, k\, = \, \infty$ as singularities, even for the anisotropic model.
However, the anisotropic $\, \chi^{(n)}$'s are sums of an {\em infinite} 
number of form factors.
One cannot try to deduce the singular points of 
these {\em infinite} sums $\, \chi^{(n)}$'s from the 
singular points of the form factors.
The  $\, \chi^{(n)}$'s are, in fact, quite
 involved ``composite'' quantities with no simple 
combinatorics interpretation (like being the sum over graphs of a certain type).
It is worth noting that exploring all the algebraic singular curves for all the
$\, \chi^{(n)}$'s, condition $\, k\, = \, 1$ always occurs for all the $\, \chi^{(n)}$'s.
  
The previous results provide a quite interesting 
insight on the ``true mathematical
and physical'' nature of the  $\, \chi^{(n)}$'s: they are quite 
involved ``composite'' quantities,
their singularities being drastically different from the 
ones of the $\, C^{(n)}(N, \, M)$ form factors~\cite{Holonomy,Assis}.

In the isotropic case, strong evidence has been 
given~\cite{Experimental,SingNfold,ongp,nic1,nic2} 
that the full susceptibility  $\, \chi$ has a 
{\em natural boundary} corresponding to the accumulation of 
singular points on the $|k| \, = \, 1$ unit circle, thus {\em discarding} 
a common wisdom that ``of course'' the singularities of the partition
function are the {\em same as the singularities of the full susceptibility}.

By analogy with the situation encountered in the isotropic case, 
we are going to have an accumulation of
singular curves densifying the whole parameter space (two 
complex variables $\, s_1$ and $\, s_2$,
i.e. four real variables). The equivalent of the unit circle is 
now, a codimension-zero manifold 
in the four real variables  parameter space, which disentangles 
two codimension-zero domains in the
parameter space. Is it the singular locus for the full 
anisotropic susceptibility $\, \chi$ ? 
Do we have here a generalization of the concept of natural boundary for several 
complex variables ?  If the answer to the question of the location
of the singularities of non-holonomic functions seems to be {\em dependent 
of the decomposition of the non-holonomic function in infinite sums 
of holonomic functions}, is it simply well-defined ? 

All we can reasonably say is that, probably, and in the same way 
as in the isotropic case, the double series for the $\,  \chi^{(n)}$'s 
are not singular in one domain (the equivalent of the inside of the unit circle), 
and one probably has the same result for the full
anisotropic susceptibility $\, \chi$.

\subsection{Anisotropic models: $\, n$-fold integrals of several complex variables}
\label{comments2}

In the anisotropic case, the $\, \chi^{(n)}$'s  are 
 $\, n$-fold integrals of several complex variables. After
 Kashiwara and Kawai~\cite{Kawai}, we 
do know that these ``functions'' of several complex variables are {\em holonomic}.
Let us restrict to the case, we often encounter in physics, where the integrand
is an {\em algebraic function} of these several complex variables (and of the 
integration variables).
In contrast with the one complex variable case, the holonomic character, 
here, corresponds to 
an extremely rich structure: the solutions of the {\em over-determined} 
system of linear PDEs correspond to a {\em finite} set of solutions 
(for one complex variable this is obvious), and the singularities, 
which are no longer points but manifolds, are {\em fixed algebraic varieties}
(for one complex variable this is obvious). Furthermore these operators 
are globally nilpotent (the holonomic functions can,
in this  ``Derived from Geometry'' framework~\cite{CalabiYauIsing1,Andre6},
be interpreted as ``Periods''
of an algebraic variety closely related to the integrand). We have many other 
remarkable properties. For instance, the operators are often (always ?) homomorphic 
to their formal adjoint (this is related to the occurrence of
selected differential Galois groups).
All these remarkable properties correspond to a differential algebra description
of these structures. Finally, we have also other properties 
of  more {\em arithmetic}
and {\em algebraic geometry} nature.
The series expansions of these holonomic functions are
 often {\em globally bounded}~\cite{Christol},
which means that they can be recast (after rescaling)
 into series expansions of several variables 
with {\em integer} coefficients.
This raises the question of the ``{\em modularity}'' 
in these problems~\cite{SP4,LianYau}.
Along this  ``modularity'' line, beyond the occurrence 
of many {\em modular forms}~\cite{CalabiYauIsing1,CalabiYauIsing},
we also see the emergence of {\em Calabi-Yau ODEs}. 
From a  differential algebra perspective, the emergence of Calabi-Yau 
structures~\cite{Calabi2} 
is not clear. In some integrability framework, the argument that Calabi-Yau 
manifolds are, after K3 surfaces, the ``next'' generalization of elliptic 
curves, remains an insufficient and much too general argument.

Let us inject, beyond the differential algebra description
of these structures, some {\em birational}
algebraic geometry ideas. In lattice statistical mechanics, the models defined 
by local Boltzmann weights depending on several complex variables,
are known to have, generically, an {\em infinite set of  birational symmetries} 
generated by the combination of the so-called {\em inversion 
relations}~\cite{inversion1,inversion2}. 

It has been shown that  $\, n$-fold integrals like the $\, \chi^{(n)}$'s
of the Ising model present some nice inversion relation functional equations
in the {\em anisotropic case}~\cite{inversion3} (several complex variables):
\begin{eqnarray}
\chi^{(n)}(K_1, \, K_2) \,\, \, = \, \,\, \, 
 \chi^{(n)}\Bigl(K_1, \, K_2\, + \, i \, {{\pi} \over {2}}\Bigr),  
\end{eqnarray}
inherited from the same inversion relation functional equation
on the full anisotropic susceptibility. 

Since the previous ideas underline the crucial role 
of the integrand of the $\, n$-fold integrals as the 
algebraic variety from which ``everything'', in principle, 
can be deduced~\cite{SingNfold,CalabiYauIsing1, Andre6}, 
it is interesting to see if this integrand, itself, is not going to be invariant 
(resp. covariant) by these  birational involutions (and, thus, by the composition 
of these birational involutions) when we keep the integration variables fixed.
One can verify that this is actually the case for the integrand of the 
anisotropic  $\, \chi^{(n)}$'s of the Ising model.

Unfortunately, the group of {\em birational transformations} of the Ising model 
is a finite set of  transformations. However, for generic models, one can easily 
imagine to be in a situation where the integrand of the $\, n$-fold integrals
of {\em several complex variables} emerging in these models, will be invariant 
(resp. covariant) by an 
{\em infinite set of birational transformations}~\cite{Automorphism}. 

We will thus have a natural emergence (in lattice statistical mechanics) 
of {\em algebraic varieties with an infinite set of
 birational symmetries}~\cite{Automorphism}. 
These algebraic varieties have zero canonical class, {\em Kodaira dimension zero}.
We, now, {\em understand the emergence of Calabi-Yau manifolds in these problems}: 
Abelian varieties and Calabi-Yau manifolds (in dimension one, elliptic curves; 
in dimension two, complex tori and K3 surfaces) have 
{\em Kodaira dimension zero}\footnote[1]{Zero canonical class,
corresponding to admitting flat metrics and Ricci flat metrics, respectively.}.

One can expect that the singular varieties (like (\ref{discrimCalab})
or (\ref{cand})) will have to be 
invariant by the (generically infinite) set of birational transformations 
generated by the inversion relations. When the singular manifolds are 
algebraic curves, the existence of a (generically infinite) set 
of birational automorphisms for the algebraic curves implies that 
the curves are, necessarily, genus zero or one~\cite{Automorphism}. This enables to 
understand\footnote[3]{Cum grano salis: in the (free-fermion) Ising case 
the birational transformations generated by the two
inversion relations form a {\em finite set}~\cite{Comment,Sacco}, 
which allows, in principle higher genus curves. 
One must imagine the Ising model as a subcase of a larger model
with $\, n$-fold integrals, where
one would recover a (generic) infinite set of birational transformations.} 
the emergence of remarkable structures like the two-parameters family of 
elliptic curves (\ref{genus1}).  Actually this is the way
many singular varieties have been discovered on many lattice statistical 
mechanics models (see~\cite{inversionWu,inversionRollet,Coxeter,Coxeter2}).  
This birational invariance fits
quite well with the interpretation of the singular variety (\ref{cand}), as 
the discriminant of a two-parameters family of Calabi-Yau 3-folds.

\vskip .1cm 

\section{Conclusion}
\label{concl}

In the theory of critical phenomena (renormalization group, etc), 
singularities are often seen as fixed points of a ``dynamical system''
called renormalization~\cite{Renorm}, and one takes for granted, with 
a (lex parsimoniae) simplicity prejudice,
that these singularities are isolated points, or smooth manifolds 
(hopefully algebraic 
varieties~\cite{challenge,challenge2,challenge3,Three-state,Three-statebis,3-12}
if one has an integrability prejudice as well). In the theory of 
discrete dynamical systems, a totally opposite prejudice exists
like the belief in a frequent occurrence of strange attractors for the set of
 fixed points of many ``dynamical systems''. Singularity theory 
in mathematics, and in particular Arnolds's viewpoint~\cite{Arnold}, 
are a perfect illustration that the set of singular points should
actually correspond to much more involved manifolds than what is 
expected in the mainstream doxa of critical phenomena.

We have performed some kind of 
``deconstruction''\footnote[2]{Using Derrida's wording.}
of the concept of singularities in lattice statistical mechanics. The sets of 
singularities are much more complex sets of points
 than what physicists tend to believe
(see Figures 1, 2, 3, 4 of~\cite{SingNfold}).

The mathematician's viewpoint that singularities are much more complex 
than what physicists believe with their (lex parsimoniae) simplicity
optimism, is the correct viewpoint. 
On the other side, the mathematician's viewpoint that nothing 
serious and/or rigorous can be done with several complex variables
is too pessimistic: within that viewpoint, singularities are seen 
as too involved to analyze, impossible to localize (of course outside the 
hypergeometric series framework), or simply, a not well-defined concept.
Even in the case of several complex variables, many singular manifolds 
conjectured by physicists,
in particular F.Y. Wu~\cite{challenge,challenge2,challenge3},
 turned out to be true singular varieties of 
lattice models, because physicists 
are (sometimes without being fully conscious) often working with 
{\em holonomic} (D-finite) functions of several complex variables. 

Focusing on the full susceptibility $\, \chi$ of the (anisotropic) Ising model
and on the holonomic $\, \chi^{(n)}$'s, we have obtained singular manifolds of the 
linear partial differential systems of the $\, \chi^{(n)}$'s. The fact that 
these  singular manifolds {\em do depend on the spectral parameter} of this 
Yang-Baxter integrable model is a strong indication that these  $\, \chi^{(n)}$'s
are {\em highly composite} objects (even if the exact expression of these 
singular varieties remains simple enough for the first $\, \chi^{(n)}$'s). 
Furthermore, the fact that most of these  singular manifolds 
are {\em not genus-zero curves} show that the series of the anisotropic $\, \chi^{(n)}$'s, 
despite all their remarkable properties, 
{\em cannot be reduced to hypergeometric series}. 

In the case of the full susceptibility $\, \chi$ of the (anisotropic) Ising model, 
we seem to have the following situation:  among the quite large, and rich,
set of singular varieties of the linear ODEs of the $\, \chi^{(n)}$'s, there is a 
restricted set (see (\ref{location2}), (\ref{genus1})) 
of singular varieties which actually corresponds to zeroes of the (anisotropic)
partition function, and, in the same time, corresponds to 
singularities of the linear PDEs of the $\, \chi^{(n)}$'s. This set could correspond 
(by analogy with the isotropic case) to {\em singularities of the series expansions} 
of the $\, \chi^{(n)}$'s. A fundamental idea to keep in mind
is that it is crucial to make a difference between the singularities 
of the (series expansions of the) D-finite functions, and the 
singularities\footnote[5]{The singular manifolds seem
 to have, in the case of $\, n$-fold integrals of algebraic integrand,
a projective invariant interpretation as discriminant of the
 algebraic varieties associated with the integrand.
} 
of the linear partial differential systems for these functions.

It would be interesting to see if, 
inside some reasonable theoretical physics framework, 
similar results\footnote[1]{With the problem that the results
seem, at first sight, to depend on the
decomposition in an infinite sum of holonomic functions.} can also be obtained 
for other {\em non-holonomic} functions 
of {\em several} complex variables that decompose into an infinite set 
of holonomic (D-finite) functions. 

\vskip .1cm

\vskip .1cm

\vskip .1cm

{\bf Acknowledgment} 
We thank J-A. Weil for help in some formal calculations on one
PDE system. We thank A. Bostan for providing a $\, p$-curvature calculation.
We thank D. Mouhanna for useful discussions.
S.B. would like to thank the LPTMC and the CNRS for kind support.
J-M.M would like to thank F. Y. Wu for so many years of a deep collaboration
only submitted to friendship and a shared love of exact 
results in lattice statistical mechanics, far from the pollution
of the short term management by project.
This present work has been performed without any support of the ANR, 
the ERC, the MAE, or any PEPS. 

\vskip .3cm

\vskip .3cm

\appendix

\section{The nine formal solutions of the Picard-Fuchs system ``above'' 
the Calabi-Yau ODE (\ref{Batyrev1})}
\label{nineformal}

Let us find the "formal solutions" around $\, (x, \, y) \, = \, \, (0, \, 0)$,
 of the PDE system (\ref{picard})
``above'' the Calabi-Yau ODE (\ref{Batyrev1}). One plugs, in (\ref{picard}), the series
\begin{eqnarray}
\label{formallist}
\sum_{j=0}  \, \sum_{k=0}^j \, \, \,  {\cal H}_{j,k} (x,y) \cdot \, \ln(x)^k \, \ln(y)^{j-k}, 
\end{eqnarray}
where ${\cal H}_{j,k} (x,y)$ are series in $x$ and $y$
and solves the system term by term. Collecting on the
non fixed coefficients, one finds $\,S_0\,=\, H_0(x,y)$ and
\begin{eqnarray}
\label{list1}
\hspace{-0.9in}&&S_1 =\,  H_0(x, \, y) \cdot \ln(x) \, + \, H_1(x, \, y), 
\qquad \quad \quad S_2 = \, H_0(x, \, y) \cdot \ln(y) \, + \, H_1(y, \, x),  
\nonumber \\
\hspace{-0.9in}&&S_3 = \, H_0(x, \, y) \cdot \ln(x)^2 \, 
+ 2\, H_1(x, \, y) \cdot \ln(x) \, + \, H_2(x, \, y),
 \nonumber \\
\hspace{-0.9in}&&S_4 =\,  H_0(x, \, y) \cdot \ln(y)^2 \, 
+ 2\, H_1(y, \, x) \cdot \ln(y) \, + \, H_2(y, \, x),
 \nonumber \\
\hspace{-0.9in}&&S_5 = \, H_0(x, \, y) \cdot \ln(x) \cdot \ln(y) \, 
+ \, H_1(y, \, x) \cdot \ln(x) \, + \, H_1(x, \, y) \cdot \ln(y)  \, + \, H_3(x, \, y),
  \nonumber \\
\hspace{-0.9in}&&S_6 = \, H_0(x, \, y) \cdot \ln(x)^2 \cdot \ln(y) \,
 + \, 2 H_1(x, \, y) \cdot \ln(x) \cdot \ln(y) \,
 + \, H_1(y, \, x) \cdot \ln(x)^2 \nonumber \\
\hspace{-0.9in}&& \qquad \, + \, 2 H_3(x, \, y) \cdot \ln(x) \, 
+ \, H_2(x, \, y) \cdot \ln(y) \, + \, H_4(x, \, y),
  \nonumber \\
\hspace{-0.9in}&&S_7 =\,  H_0(x, \, y) \cdot \ln(x) \cdot \ln(y)^2 \,
 + \, 2 H_1(y, \, x) \cdot \ln(x) \cdot \ln(y) \, + \, H_1(x, \, y) \cdot \ln(y)^2 
\nonumber \\
\hspace{-0.9in}&& \, + \, 2 H_3(x, \, y) \cdot \ln(y) \, + \, H_2(y, \, x) \cdot \ln(x) \,
 + \, H_4(y, \, x),
  \nonumber \\
\hspace{-0.9in}&&S_8 = \, H_0(x, \, y) \cdot \ln(x)^2 \cdot \ln(y)^2 \, 
+ \, 2 H_1(y, \, x) \cdot \ln(x)^2 \cdot \ln(y) \,
+ \, 2 H_1(x, \, y) \cdot \ln(x) \cdot \ln(y)^2 \,
  \nonumber \\
\hspace{-0.9in}&&\qquad + \, 4 H_3(x, \, y) \cdot \ln(x) \cdot \ln(y)  \,
 + \, H_2(y, \, x) \cdot \ln(x)^2 \,+ \, H_2(x, \, y) \cdot \ln(y)^2 \, 
\nonumber \\
\hspace{-0.9in}&&\qquad + \, 2 H_4(y, \, x) \cdot \ln(x) \,
+ \, 2 H_4(x, \, y) \cdot \ln(y) \, + \, H_5(x, \, y), 
\end{eqnarray}
where  (only the first terms of the series are given)
\begin{eqnarray}
\fl \qquad H_0(x,y)\, = \,\, \,\,  1\, \,\, + 6\,(x+y) \, \, \,
+ (90  \,(x^2+y^2) \,+720 \,xy) \, \, \, + \, \, \cdots, 
\nonumber \\
\fl \qquad  H_1(x,y) \,=\, \,\, (15\,x\,+33\,y) \,\,
 +\Bigl({513 \over 2} x^2 +3132\,xy\,+{1323 \over 2} \, y^2\Bigr) 
\,\,\, + \, \, \cdots,
 \nonumber  \\
\fl \qquad  H_2(x,y) \,=\,\,\,   (108\,y\,-18\,x) \,\, 
 -\left({279 \over 2} x^2 \,-6120 xy \,-3654 y^2 \right) 
 \,\,\, + \, \, \cdots,
 \nonumber  \\
\fl \qquad  H_3(x,y) \,=\, \,\,  9 \cdot \, (x+y) \, \,\,
 +\left({2709 \over 4} x^2 \, +3960 \, xy + {2709 \over 4}\, y^2 \right) 
 \, \,\, + \, \, \cdots, 
\nonumber  \\
\fl \qquad  H_4(x,y) \,=\,\,\,
 - \, (90\, x \, +162\, y) \, \, \, 
- \left({{8505} \over {4}} \, x^2  \,+11178 x y  \,+ {{6237} \over {4}} \, y^2 \right)
  \, \,\,\, + \, \, \cdots, 
\nonumber  \\
\fl \qquad  H_5(x,y) \,=\, \,\, 324  \cdot \, (x\, + y) \,\,
 -\left({{14931} \over{4}} (x^2\,+y^2) \, -6912 x y \right)
 \,\, \,\, + \, \, \cdots
 \nonumber  
\end{eqnarray}

There are {\em nine solutions} for the system (\ref{picard}). One notes 
that $H_0$, $H_3$ and $H_5$ are symmetric in $x$, $y$, while $H_1$, $H_2$ and $H_4$ 
are not symmetric in $x$, $y$. For the formal solutions, $S_0$, $S_5$ and $S_8$ 
are symmetric in $x$, $y$, and the six others are pairwise symmetric.
These nine independent formal solutions are  solutions of 
the PDE system (\ref{picard}), and thus of 
the order-nine  differential operator $\, \tilde{\Omega}_x$  and 
its $\, (x, \, y)$-symmetric $\, \tilde{\Omega}_y$. 

Note however, that the linear differential operator $\, \tilde{\Omega}_x$ has been 
constructed from the PDE system (\ref{picard}) and factorizes as written 
in (\ref{factoL9L6}), it, then, might be that $\, H_0(x,y)$ is a solution 
of only the right factor operator $\, L_6(x,y)$. Indeed, plugging a series
\begin{eqnarray}
\label{genS}
\sum_{n,m}\, c_{n,m} \cdot \, x^n \, y^m, \qquad \qquad c_{n,m}\,\, = \,\,\, c_{m,n}, 
\end{eqnarray}
into $\, L_6(x,y)$ and solving term by term, one obtains (up to the overall $c_{0,0}$),
 the double hypergeometric series $\, H_0(x, \, y)$. The solutions of  $\, L_6(x,y)$
can be expressed in terms of the previous formal solutions (\ref{list1}):
\begin{eqnarray}
\label{solL6formal}
\hspace{-0.7in}S_0, \,\quad \quad  S_1, \,\quad \quad  S_2, \,\quad \quad 
 S_3\, -\, S_4, \, \quad \quad  \, S_5 \, +\, {{S_4} \over {2}}, \,\quad  \quad 
S_6 \, +\, S_7.
\end{eqnarray}

\section{Factorization (\ref{factoL9L6}) of the order-nine 
operator  $\, {\tilde \Omega}_x$}
\label{L6}

The order-nine operator $\, {\tilde \Omega}_x$ of subsection (\ref{otherpicard})
factorizes (see (\ref{factoL9L6}))
into three order-one operators 
and the order-six operator $\, L_6(x, \, y)$:
\begin{eqnarray}
L_6(x, \, y) \,  \, \, = \, \, \,  \,\,  \,
{{1} \over {p_6(x, \, y)}}  \cdot \, \sum_{n=0}^6 \, p_n(x, \, y) \cdot D_x^n, 
\end{eqnarray}
The three order-one operators are encoded by three rational functions of $\, x$ 
and $\, y$, 
namely $\, \tilde{r}_1(x,\, y)$, $\, \tilde{r}_2(x,\, y)$ 
and $\, \tilde{r}_3(x,\, y)$. These polynomials factorize (see (\ref{qn}))
and thus the $\, \tilde{r}_i(x,\, y)$'s reduce to 
the expressions of four polynomials with integer coefficients
$\, {\cal P}_9(x, \, y)$, $\, {\cal P}_6(x, \, y)$,  $\, q_1$ and $\, q_2$, 
where  $\, {\cal P}_9(x, \, y)$ is the polynomial of the apparent singularities
of the order-nine operator $\, {\tilde \Omega}_x$, and where  
$\, {\cal P}_6(x, \, y)$ is the polynomial of the apparent singularities
of the order-six operator $\, L_6(x, \, y)$.

These polynomials read:
\begin{eqnarray}
\label{calP9}
\hspace{-0.9in}&&{\cal P}_9(x, \, y)  \,  \, \, = \, \, \, \, 
2^4 \cdot 3^{18} \cdot \,{x}^{6} \, \,  \,
- 2 \cdot 3^{16} \, \cdot \, (31951 +1602072\,y) \cdot \,  {x}^{5} \, 
\nonumber \\
\hspace{-0.9in}&& \qquad
+ 3^{13} \, \cdot \,  (14397329+ 913784868\,y +17712588816\,{y}^{2}) \cdot \,  {x}^{4}\, 
\nonumber \\
\hspace{-0.9in}&&\qquad + 3^9 \,\cdot \, 
(2986814425 +60616383939\,y -1350750590172\,{y}^{2}
\nonumber \\
\hspace{-0.9in}&& \qquad \qquad \qquad 
-24695209500192\,{y}^{3}) \cdot \,  {x}^{3} 
\nonumber \\
\hspace{-0.9in}&&\qquad
\, + 3^7\,\cdot \, (5310925151 -333452529387\,y -14254789072275\,{y}^{2}
\nonumber \\
\hspace{-0.9in}&&\qquad \qquad \qquad
 +241096254564492\,{y}^{3} +7702353325801296\,{y}^{4})
 \cdot \,  {x}^{2} \, 
\nonumber \\
\hspace{-0.9in}&&\qquad -81\, \cdot \, (27\,y-1) \cdot \, 
 \, (39319888296092688\,{y}^{4}
+122020942792986\,{y}^{3} \nonumber \\
\hspace{-0.9in}&& \qquad  \qquad  \qquad   -111685613173821\,{y}^{2}+
22118310900\,y +86524357339) \cdot \, x \, 
\nonumber \\
\hspace{-0.9in}&&\qquad 
+ 2^4 \cdot 5^3 \, \cdot \, (10827\,y+364) ^{3} \cdot \, (27\,y-1)^{3}, 
\end{eqnarray}
\begin{eqnarray}
\label{calP6}
\hspace{-0.9in}&& {\cal P}_6(x, \, y) \, = \, \, 
\, 387420489 \cdot \, ({x}^{2}-142\,xy+343\,{y}^{2}) \cdot \,  (x+y)^{4} \,
\nonumber \\
\hspace{-0.9in}&&  \qquad 
-43046721 \cdot \, (x+y) \cdot \, 
(89\,{x}^{4}-196\,{y}^{4}-823\,x{y}^{3}+13287\,{x}^{2}{y}^{2}-3493\,{x}^{3}y) \, 
\nonumber \\
\hspace{-0.9in}&& \qquad 
+1594323 \cdot \, (3482\,{x}^{4} +662\,x{y}^{3} +2972\,{x}^{3}y
-427\,{y}^{4} +25365\,{x}^{2}{y}^{2})
\nonumber \\
\hspace{-0.9in}&& \qquad 
\, +19683 \cdot \, (33307\,{x}^{3} -1784\,{y}^{3} -14487\,x{y}^{2} +44904\,{x}^{2}y)
 \\
\hspace{-0.9in}&& \qquad 
\, -2187 \cdot \, (27394\,{x}^{2}-88\,xy-671\,{y}^{2})
 \quad 
\, +162 \cdot \, (1325\,x+242\,y) \,  \,  -1331, \nonumber
\end{eqnarray}
\begin{eqnarray}
\hspace{-0.9in}&&q_1 \, = \, \,\, \,
4 \cdot 3^{18}  \cdot \, ({x}^{6}+113061462\,x{y}^{5}
 +4560\,{x}^{5}y -8876482\,{x}^{3}{y}^{3}
+284847\,{x}^{4}{y}^{2}
\nonumber \\
\hspace{-0.9in}&& \qquad \qquad 
-52726107\,{x}^{2}{y}^{4}+28140175\,{y}^{6}) \, 
\nonumber \\
\hspace{-0.9in}&&\qquad 
+3^{16}  \cdot \, 
(4108\,{x}^{5} -11112875\,{x}^{3}{y}^{2} +587276\,{x}^{4}y 
-105291883\,x{y}^{4}
\nonumber \\
\hspace{-0.9in}&& \qquad  \qquad \qquad 
+14516200\,{y}^{5}-4491914\,{x}^{2}{y}^{3}) 
\nonumber \\
\hspace{-0.9in}&&\qquad 
\, +3^{13}  \cdot \, (198311\,{x}^{4} \, -370624786\,x{y}^{3} \,
 +6765614\,{x}^{3}y  
\nonumber \\
\hspace{-0.9in}&& \qquad \qquad \qquad 
-130714000\,{y}^{4}+116112144\,{x}^{2}{y}^{2}) \, 
\nonumber \\
\hspace{-0.9in}&&\qquad 
+3^9 \cdot \, 
(18879841\,{x}^{3}-64727000\,{y}^{3} \, +773936148\,x{y}^{2} \,
+17519934\,{x}^{2}y) 
 \nonumber \\
\hspace{-0.9in}&&\qquad 
-3^7 \cdot \, (45403057\,{x}^{2}-221205178\,xy -141045500\,{y}^{2})
 \nonumber \\
\hspace{-0.9in}&&\qquad 
-567 \cdot \, (22002263\,x-1112800\,y) \,\,  \,  -145745600, 
\nonumber 
\end{eqnarray}
\begin{eqnarray}
\hspace{-0.9in}&&q_2 \, = \, \, \,\,
 774840978 \cdot \, ({x}^{6}\, -841926\,x{y}^{5} \,-462\,{x}^{5}y
\, -341728\,{x}^{3}{y}^{3} \, +32721\,{x}^{4}{y}^{2}
\nonumber \\
\hspace{-0.9in}&&\quad \qquad 
+810681\,{x}^{2}{y}^{4}+98245\,{y}^{6}) \, 
\nonumber \\
\hspace{-0.9in}&&\quad \qquad 
-43046721 \cdot \,(223\,{x}^{5} +54121\,{x}^{3}{y}^{2}
 -47245\,{x}^{4}y -613336\,x{y}^{4}
\nonumber \\
\hspace{-0.9in}&&\quad \qquad 
-68810\,{y}^{5}\, +20707\,{x}^{2}{y}^{3}) \, 
\nonumber \\
\hspace{-0.9in}&&\quad \qquad 
+1594323 \cdot \, (22489\,{x}^{4}\, +1358236\,x{y}^{3}\, 
+304861\,{x}^{3}y\,
\nonumber \\
\hspace{-0.9in}&&\quad \qquad 
 -250820\,{y}^{4} \, -1645923\,{x}^{2}{y}^{2}) \,
 \nonumber \\
\hspace{-0.9in}&&\quad \qquad 
+19683\, (415049\,{x}^{3}\, -505660\,{y}^{3}\, -4725138\,x{y}^{2}\, 
+65103\,{x}^{2}y) \, 
 \nonumber \\
\hspace{-0.9in}&&\quad \qquad 
+10935 \cdot \, (229157\,{x}^{2}\, -163880\,xy\, +68006\,{y}^{2})
\nonumber \\
\hspace{-0.9in}&&\quad \qquad 
+162 \cdot \, (492079\,x+45925\,y) \,\, \,-440440. 
\nonumber 
\end{eqnarray}

\section{Alternative linear differential operator for 
the double hypergeometric series}
\label{further2}

Recalling the double hypergeometric series (\ref{H2}), 
 $\, H_0(x, \, c\, x)$ is solution of an order-six 
$\, c$-dependent linear differential operator
\begin{eqnarray}
\hspace{-0.9in}&&W_6 \,  \, \, = \, \, \,   \,\, 
 (1 \, +162 \cdot (c+1) \cdot x)\,  \times 
\nonumber \\
\hspace{-0.9in}&& \quad  \quad (1 \,  \, -81 \cdot (c+1) \cdot x \,
+2187 \cdot (c^2-7\,c+1) \cdot x^2 \,\,
 -19683 \cdot (c+1)^3 \cdot x^3\,) \cdot x^4 \cdot D_x^6
 \nonumber \\
\hspace{-0.9in}&&  \quad  \quad \quad \,  \, \, + \,\,  \, \cdots 
\end{eqnarray}

In the $\, c \, = \, 1$ limit, 
this order-six operator becomes the direct sum 
of the order-two linear differential operator 
\begin{eqnarray}
\label{secorderbis}
\theta^2 \,\, -\, 3\, x \cdot  (3\, \theta \, +\, 1) \cdot   (3\, \theta \, +\,2), 
\nonumber 
\end{eqnarray}
with the hypergeometric function solution 
\begin{eqnarray}
 _2F_1\Bigl([{{1} \over {3}}, \,{{2} \over {3}}], [1]; \, -27 \, x\Bigr), 
\end{eqnarray}
and of  the order-four Calabi-Yau ODE (\ref{Batyrev1}),
with the analytic  solution (\ref{solBat1}), which can be written as
the Hadamard product~\cite{Hadamard}:
\begin{eqnarray}
\label{hadprodbat1}
\hspace{-0.5in} _2F_1\Bigl([{{1} \over {3}}, \,{{2} \over {3}}], [1];
 \, -27 \, x\Bigr) \, \star \, 
\Bigl( {{1 } \over {1\, -4 \, x}}\,  \cdot \,
 _2F_1\Bigl([{{1} \over {3}}, \, {{2} \over {3}}], \, [1]; \, 
-{{27 \cdot x } \over {(1\, -4 \, x )^3 }}  \Bigr) \Bigr).
\nonumber 
\end{eqnarray}

In the (less natural)  $\, c \, = \, 0$ limit, 
this order-six  linear differential operator is the 
product of homomorphic operators:
\begin{eqnarray}
W_6(c=0) \,\,\,  \,= \, \, \,\, \, \, 
 N_2 \cdot M_2 \cdot L_2,
\end{eqnarray}
where $\, L_2$ has the hypergeometric function solution
\begin{eqnarray}
 _2F_1\Bigl([{{1} \over {3}}, \,{{2} \over {3}}], [1]; \, 27 \, x\Bigr). 
\end{eqnarray}
In the $\, c \, \rightarrow \, \infty$ limit, this order-six operator 
degenerates into the  direct sum:
\begin{eqnarray}
\hspace{-0.8in}(3 \cdot \theta \, + \, 1) \oplus (3 \cdot \theta \, + \, 2)
 \oplus (3 \cdot \theta \, + \, 4) \oplus
 (3 \cdot \theta \, + \, 5) \oplus 
(3 \cdot \theta \, + \, 7) \oplus (3 \cdot \theta \, + \, 8).
\nonumber  
\end{eqnarray}

\section{Another series of two complex variables}
\label{Kampe}

\subsection{Double hypergeometric series}
\label{doublehyp}

Without the factor 64, the results for (\ref{poch}) 
in subsection (\ref{picardmorebat2}) correspond to  
 the double hypergeometric series
\begin{eqnarray}
\label{Kampdef}
\hspace{-0.3in}K(x, \, y) 
\, \,  \, = \, \, \, \,  \,  \,  \, 
\sum_{n\, = \, 0}^{\infty} \, \sum_{m\, = \, 0}^{\infty}  
\, {{(\alpha)_n^3 \cdot (\beta)_m^3 \cdot (\beta')_{m+n} } \over {
(\gamma)_{m+n}^3 \, \, n! \, \, m! }} \cdot x^n \cdot y^m, 
\nonumber 
\end{eqnarray}
where $\, (\alpha)_n$ is the usual Pochhammer symbol. 
The double hypergeometric series 
$\, K(x, \,  y)$ is a {\em Kamp\'e-de-F\'eriet} 
function~\cite{Gaveau,Gaveau2,Kamp,Appell}
\begin{eqnarray}
\label{Kamp}
\hspace{-0.5in}
F^{1,3,3}_{3,0,0} ( [\beta'],\, [\alpha,\,\alpha,\,\alpha],\,
 [\beta,\,\beta,\,\beta]; [\gamma,\, \gamma,\, \gamma],-,-; \,x,\,y).
\end{eqnarray}

The singularity varieties of (\ref{Kampdef}) are 
{\em independent} of the parameters 
$\, \alpha$, $\, \beta$,   $\, \beta'$,   $\, \gamma$, and are
$\,\, x \cdot \,  (1-x)\cdot \,  (1-y) \cdot \,  (y-x) \, \, = \, \, \, 0$, 
together with 
\begin{eqnarray}
\label{singKamp}
\hspace{-0.1in}
 y^2\,  x^2 \,  \,\,  \, -2\,  x\,  y \cdot \,  (y+x) \, \,\,  +(x-y)^2
 \, \,\,\,  = \, \, \,\,\,  \, 0, 
\end{eqnarray}
in agreement, in the $\,  \alpha \, =\, \beta \, =\, \beta' \, = \, 1/2$, 
  $\, \gamma \, = \, 1$ limit, 
 with (\ref{tildeS2xy}), taking into account the rescaling
$ \, \, (x, \, y) \, \rightarrow \,  (64 \, x, \, 64 \, y)$.

\vskip .1cm 

\subsection{Other double hypergeometric series}
\label{otherdoublehyp}

Introducing the other double hypergeometric series
\begin{eqnarray}
\label{Kampdef2}
\hspace{-0.3in}K_2(x, \, y) 
\, \, \,  \, = \,  \,\, \, \,  \, \sum_{n\, = \, 0}^{\infty} \, \sum_{m\, = \, 0}^{\infty}  
\, {{(\alpha)_n^M \cdot (\beta)_m^M \cdot (\beta')_{m+n} } \over {
(\gamma)_{m+n}^M \, \, n! \, \, m! }} \cdot x^n \cdot y^m.
\end{eqnarray}
It is  also a {\em Kamp\'e-de-F\'eriet} function~\cite{Kamp,Appell,Gaveau,Gaveau2}
\begin{eqnarray}
\label{Kampother}
\hspace{-0.5in}
F^{1,M,M}_{M,0,0} ( [\beta'],\, [\alpha,\,\cdots ,\,\alpha],\,
 [\beta,\,\cdots,\,\beta]; [\gamma,\,\cdots,\, \gamma],-,-; \,x,\,y).
\end{eqnarray}
Let us restrict, in the following, to 
$\, \alpha \, =  \, \,   \beta \, =  \,  \, \beta' \, = \, 1/2$
 and  $\,\gamma \, = \, 1$. 

The singularity varieties of the PDE system are actually different from 
(\ref{singKamp}) and depend on $\, M$. For $\, M\, = \, \, 2$ and 
$\, M\, = \, 4$, they read respectively:
\begin{eqnarray}
\label{M4}
\hspace{-0.6in}(x+y)^2\, - \, x^2\, y^2\,\,\,  = \,\,\,\,\,0,
 \qquad \quad 
(x+y -\, x\, y)^3\, + \, 27 \, x^2\, y^2\,\,\,  = \,\,\,\,\,0.
\end{eqnarray}
More generally, for $\,M$ an even integer, besides the conditions 
$\, x \cdot \,  (1-x) \cdot \,  (1-y) \, = \, \, 0$,
the singular manifold   reads an algebraic curve 
of parametrization 
\begin{eqnarray}
\label{param1}
\hspace{-0.3in}x \, \, = \, \, \, t^{M-1}, 
\qquad \quad y \, \, = \, \, \, \Bigl( {{- \, t} \over {1 \, -t}} \Bigr)^{M-1}, 
\end{eqnarray}
or equivalently 
\begin{eqnarray}
\hspace{-0.3in}x \, \, = \, \, \, \Bigl({{1} \over {2}} \, +v\Bigr)^{1-M}, 
\qquad \quad y \, \, = \, \, \, \Bigl({{1} \over {2}} \, -v\Bigr)^{1-M}, 
\end{eqnarray}
that can be thought as a ``Fermat-like'' curve:
\begin{eqnarray}
\hspace{-0.3in}x^{{{1} \over {1-M}}} \, \, \,  + \, y^{{{1} \over {1-M}}} 
\,  \, \, \,  =\, \, \,   \, 1. 
\end{eqnarray}

For $\, M\, = \, \, 3$, we have (\ref{singKamp})  
and for $\, M\, = \, \, 5$,
we have (besides the conditions 
$\, x \cdot \,  (1-x)\cdot \,  (1-y) \cdot \,  (y-x)\, = \, \, 0$)
the singular variety
\begin{eqnarray}
\hspace{-0.7in}&&(x+y\, +x\, y)^4\, \, \, \, 
 -136\, x^2\, y^2 \cdot \, (x+y\, +x\, y)\, \,
\,  \,  -8\, x\, y \cdot \, (x+1+y) \, (x^2+y^2)\, \,
 \nonumber \\
\hspace{-0.7in}&& \qquad \qquad \quad \, 
  -8\, x^2\, y^2 \cdot \, (x+y)\, (x\, y-1)
 \, \,\,\,\, = \,\, \,\,\, \, 0. 
\end{eqnarray}
More generally, for $\,M$  an odd integer, besides the conditions 
$\, x \cdot \,  (1-x)\cdot \,  (1-y) \cdot \,  (y-x)\, = \, \, 0$,
the singular manifold  reads an algebraic curve of parametrization 
\begin{eqnarray}
\label{param2}
\hspace{-0.3in}x \, \, = \, \, \, t^{M-1}, 
\qquad \quad y \, \, = \, \, \, \Bigl({{-\, t} \over {1 \, -t}} \Bigr)^{M-1}, 
\end{eqnarray}
or equivalently 
\begin{eqnarray}
\hspace{-0.3in}x \, \, = \, \, \, \Bigl(-{{1} \over {2}} \, +v\Bigr)^{1-M}, 
\qquad  \quad y \, \, = \, \, \, \Bigl(-{{1} \over {2}} \, -v\Bigr)^{1-M}, 
\end{eqnarray}
that can be thought as a ``Fermat-like'' curve:
\begin{eqnarray}
\hspace{-0.3in}x^{{{1} \over {1-M}}} \, \, \,  + \, y^{{{1} \over {1-M}}} 
\, \, +\, \, 1 \, \,  \,  \, = \,  \, \,  \, 0. 
\end{eqnarray}

\subsection{Differential operators restricted to singular varieties}
\label{singoper}

Let us restrict to the {\em singular variety}
 (\ref{singKamp}) for $\, M\, = \, 3$,
using the rational parametrization (\ref{param2}), that is 
$\, (x, \, y) \, = \, \, (t^2, \, (t/(1-t))^2)$. The double 
series expansion (\ref{Kampdef2})
becomes a series expansion in the $\, t$ variable which is solution of the 
order-four linear differential operator  ($D_t \, = \, d/dt$):
\begin{eqnarray}
\label{calC4}
\hspace{-0.7in}&&{\cal C}_4 \, \, = \, \, \, t^3 \cdot \, (t-1)
  \, (2\,t+1)  \, (t+2)  \, ({t}^{2}+t+1)^{2}
 \, \left( t+1 \right)^{4} \cdot \,  D_t^{4} \, 
 \nonumber \\ 
\hspace{-0.7in}&& \qquad \, +2\, t^2 \cdot \, ({t}^{2}+t+1)  \, \cdot \  \,
 (t+1)^{3}\,\cdot \,  c_3(t) \cdot \,  D_t^3
\, \, \,\,  +t \cdot \,  (t+1)^{2} \cdot \, c_2(t) \cdot \,  D_t^2
\nonumber \\ 
\hspace{-0.7in}&& \qquad +2\, (t+1) \cdot 
  \, c_1(t) \cdot \,  D_t
\, \,\,\, \,  +2\,\,t \,\cdot \, (t+2)  \, ({t}^{2}+t+1)^{4}, 
\end{eqnarray}
where
\begin{eqnarray}
\hspace{-0.9in}&&c_3(t) \, = \, \, \,
 10\,{t}^{6} +32\,{t}^{5}+39\,{t}^{4}+20\,{t}^{3}-17\,{t}^{2}-24\,t-6, 
 \nonumber \\ 
\hspace{-0.9in}&&c_2(t) \, = \, \, \,50\,{t}^{9}+243\,{t}^{8}
+588\,{t}^{7}+903\,{t}^{6}+885\,{t}^{5}
+501\,{t}^{4}+33\,{t}^{3}-174\,{t}^{2}-99\,t-14,
 \nonumber \\ 
\hspace{-0.9in}&&c_1(t) \, = \, \, \,15\,{t}^{10}+82\,{t}^{9}
+228\,{t}^{8}+411\,{t}^{7}+531\,{t}^{6}+513\,{t}^{5}+333\,{t}^{4}
 \nonumber \\ 
\hspace{-0.9in}&& \qquad \quad \quad \quad 
+99\,{t}^{3}-12\,{t}^{2} -12\,t-1, 
\nonumber 
\end{eqnarray}
This ``critical'' order-four operator ${\cal C}_4$ 
is such that its {\em exterior square}
 is a linear differential operator of order 
{\em five} (and not six as it should be
for a generic order-four operator). This condition 
that the exterior square  is of order five
 is called the ``Calabi-Yau 
condition'':
it is one of the conditions defining Calabi-Yau 
ODEs~\cite{TablesCalabi,Almkvist1,Almkvist2,Almkvist3}. 
Related to this exterior square condition one 
also has the property that this order-four operator 
${\cal C}_4$ is homomorphic to its adjoint, up to a 
conjugaison by the polynomial $\,(x+1)^3\, (x^2+x+1)^3$. 

Note that the limit $\, y \, = \, x$, yielding to 
the Calabi-Yau operator (\ref{defBatyrev2}) 
(also such that its {\em exterior square}
 is a linear differential operator of order {\em five}), is 
{\em actually a singular limit}
 of the Picard-Fuchs system. 

Similarly, let us restrict to the {\em singular variety} 
(\ref{M4}) for $\, M\, = \, 2$,
using the rational parametrization (\ref{param1}), namely 
$\, (x, \, y) \, = \, \, (t, \, -t/(1-t))$.  
The double series expansion (\ref{Kampdef2})
becomes a series expansion in the $\, t$ variable which is solution of the 
order-three linear differential operator  ($D_t \, = \, d/dt$):
\begin{eqnarray}
\hspace{-0.7in}&&{\cal C}_3 \, \, = \, \, \, \,  \,D_t^3 \, \, \,  
 + \, {{3} \over {2}} \cdot \,{\frac { (3\,t-2) }{t \, (t-1) }} \cdot D_t^2
\,\,  \,  + \, {{1} \over {4}}\cdot 
\,{\frac { 13\,{t}^{2}-16\,t+4 }{ (t-1)^{2} \cdot \, t^{2}}} \cdot D_t 
\,  \,  + \, {{1} \over {8}} \cdot \,{\frac {t-2}{t \cdot \, (t-1)^{3}}}.
\nonumber 
\end{eqnarray}
This ``critical'' order-three operator ${\cal C}_3$
 is such that its {\em symmetric square} is a linear differential 
operator of order {\em five} (and not six as it should be for a generic 
order-three operator). 
Related to this last property one 
also has the property that this order-three operator 
${\cal C}_3$ is homomorphic to its adjoint, up to a 
conjugaison by the rational function $\,1/x^2/(x-1)$. 

This order-three operator ${\cal C}_3$ is, in fact, 
exactly the symmetric square of 
\begin{eqnarray}
16\, t \cdot \, (t-1)^{2} \cdot D_t^{2} \, \, \,  
+8 \cdot \, (3\,t -2)  \cdot \, (t-1) \cdot \,  D_t  \,   \,\, +t, 
\end{eqnarray}
which has $\, (1-t)^{1/4} \cdot \, K(t^{1/2})$ as a solution ($K$ is the 
complete elliptic integral of the first kind).

\vskip .1cm 
Let us now restrict to the {\em singular variety}
 (\ref{M4})  for $\, M\, = \, 4$,
using the (alternative) rational parametrization 
\begin{eqnarray}
x \,\, = \, \, \,8 \, t, \qquad \quad 
y \,\, = \,\, \, - \, {{8 \, t} \over { 1 \, -\, 8 \, t}}.
\end{eqnarray}
With this parametrization the  double 
series expansion (\ref{Kampdef2})
becomes a series expansion in the $\, t$ variable 
with {\em integer coefficients}.
It is solution of an order-eight operator, its symmetric square 
is of order $\, 35$ (and not $\, 36$ as it 
should be generically\footnote[1]{Its exterior square is 
order 28 as it should for a generic order-eight operator.}).

For $\, M\, = \, \, 4$ the double series can also be 
resummed in one variable and rewritten as 
\begin{eqnarray}
\hspace{-0.9in}\sum_{m \, = \, 0}^{\infty} \, 
 {{ (2m)!^5 } \over {4^{5\, m} \cdot \, m!^{10} }} \,     \cdot \, 
_5F_4\Bigl([{{1} \over{2}}, \, {{1} \over{2}}, \, {{1} \over{2}}, \, {{1} \over{2}},
 \, m \, + \, {{1} \over{2}} ], 
\, [m+1, \,m+1, \,m+1, \,m+1 ]; x   \Bigr) \cdot y^m, \nonumber  
\end{eqnarray}
corresponding to the identity
\begin{eqnarray}
\hspace{-0.7in}{{ (2m)!^5 } \over {4^{5\, m} \cdot \, m!^{10} }} \,     \cdot \, 
\Bigl( {{(1/2)_n^4 \cdot (m+1/2)_n  } \over { n! \cdot \, (m+1)_{n}^4}} \Bigr)
\, \, \, = \, \, \, \,\,  
{{(1/2)_n^4 \cdot (1/2)_m^4 \cdot (1/2)_{m+n} } \over { n! \cdot \, m! \cdot \,(1)_{m+n}^4}}.
  \nonumber  
\end{eqnarray}
More generally one has the identities
\begin{eqnarray}
{{ (2m)!^M } \over {4^{M\, m} \cdot \, m!^{2\, M} }} \, \, \, = \, \, \, \,\,  
{{ (1/2)_m^M  } \over {m!^{M}}}, 
\end{eqnarray}
and 
\begin{eqnarray}
\hspace{-0.9in}{{ (2m)!^{M+1}} \over {4^{(M+1) \, m} \cdot \, m!^{2 \, (M+1) } }} \,     \cdot \, 
\Bigl( {{(1/2)_n^M \cdot (m+1/2)_n  } \over { n! \cdot \, (m+1)_{n}^M}} \Bigr)
\, \, \, = \, \, \, \,\,  
{{(1/2)_n^M \cdot (1/2)_m^M \cdot (1/2)_{m+n} } \over { n! \cdot \, m! \cdot \,(1)_{m+n}^M}},
  \nonumber  
\end{eqnarray}
and the alternative writing of the 
 the double series (\ref{Kampdef2}), for 
 $\, \alpha \, =  \, \,   \beta \, =  \,  \, \beta' \, = \, 1/2$
 and  $\,\gamma \, = \, 1$, as 
\begin{eqnarray}
\hspace{-0.6in}&& \sum_{m \, = \, 0}^{\infty} \,  \, 
 {{ (2m)!^{M+1} } \over {4^{ (M+1) \, m} \cdot \, m!^{2 \, (M+1) } }} \, 
\quad \times \\
\hspace{-0.6in}
&& \qquad \quad \quad 
_{M+1}F_M\Bigl([{{1} \over{2}}, \, \cdots, \, {{1} \over{2}},
 \, m \, + \, {{1} \over{2}} ], 
\, [m+1, \, \cdots, \, \,m+1 ]; x   \Bigr) \cdot y^m, 
\nonumber  
\end{eqnarray}

\vskip .1cm 

Let us now restrict to the {\em singular variety} $\, y \, = \, \, x$.
For $\, M\, = \, 4$ and  $\, M\, = \, 5$, the  double 
series expansion (\ref{Kampdef2})
becomes a series expansion in $\, x$ that is solution of an order-six 
linear differential operator. For  $\, M\, = \, 4$, 
this order-six operator is such that its symmetric square is 
of order $\, 20$ (instead of the order $\, 21$ one 
could expect generically). For  $\, M\, = \, 5$, 
this order-six operator is such that its exterior square is 
of order $\, 14$ (instead of the order $\, 15$ one could expect generically).

\section{More Picard-Fuchs systems above Calabi-Yau ODEs}
\label{morepic}

\subsection{More Picard-Fuchs system with two variables}
\label{picardmorebat5}
\vskip .1cm 

Another example is the two-variables Picard-Fuchs system 
``above'' the order-four Calabi-Yau operator (see ODE number
 18 in~\cite{TablesCalabi}))
\begin{eqnarray}
\label{Batyrev5}
&&\theta^4 \,\, -4 \, x \cdot 
(3 \, \theta^2 \, +3 \,\theta \, +1) \cdot   (2\, \theta \, +\, 1)^2
\,  \nonumber \\
&&\quad \quad  \, \,  -4 \, x^2 \cdot  
   (4\, \theta \, +\, 5) \cdot   (4\, \theta \, +\,6) \cdot
   (4\, \theta \, +\, 2) \cdot
   (4\, \theta \, +\, 3)    \\
&&\quad \quad  \quad  \, = \, \, \,
\,   (1\,-64\, x)  \cdot (1\, +16\, x) \cdot x^4   \cdot D_x^4
 \, \,\, + \, \, \cdots   \nonumber 
\end{eqnarray}
The Picard-Fuchs system corresponds to
the double series~\cite{Batyrev} 
\begin{eqnarray}
\label{defBat5def}
\hspace{-0.9in}&&\sum_{n=0}^{\infty} \,  \sum_{m=0}^{\infty} \,  
{{ (n+m)!^2 \, \, (2\, m \, +2\, n)!} \over { n!^4 \, \,m!^4 }} \cdot \, x^n \, y^m 
\,\, \, \,\,  =
\nonumber  \\
\hspace{-0.9in}&& \quad \quad \,\, = \, \, \, \sum_{m=0}^{\infty} \, 
 \, {{(2\, m)!} \over {m!^2}} \, \cdot \, 
_4F_3\Bigl([m\, +1, m\, +1, m\, +1, \,m\, +{{1} \over {2}}],
 \, [1, \, 1, \, 1]; \, 4 \, y   \Bigr)
 \cdot \, x^n \nonumber 
\end{eqnarray}
\begin{eqnarray}
\label{defBat5}
\hspace{-0.9in}&& \quad \quad \,\, = \, \, \, \,
1 \, \,\, +2 \, \cdot \, (x+\, y)\, \,+ 6 \cdot \, (x^2\, + y^2\, +\, 16\, x\, y)\, 
 +20\, (y+x)\, (x^2 \, +y^2 \, +80\, x\, y) \,
 \nonumber  \\
\hspace{-0.9in}&&\quad \quad \quad \quad \quad \, \, \, \, 
+ 70 \cdot \, ({x}^{4}+{y}^{4} \, +256\,{x}^{3}y+256\,x{y}^{3} \,
 +1296\,{x}^{2}{y}^{2}+{x}^{4})  \,\, \, + \, \, \cdots 
\end{eqnarray}  
Note that all the coefficients of odd orders in $\, x$ and $\, y$ 
factor $(x+y)$.

The  singular variety is the union of 
$\,\,\, x \, y \cdot \, (x-y)  \, = \, 0\,\,$ 
together with the $(x,\,y)$-symmetric {\em genus-zero} algebraic curve 
which reads:
\begin{eqnarray}
\label{singBat5}
\hspace{-0.6in}&&2^8 \cdot \,(x-y)^4\, \, \,  \, 
-2^8 \cdot \, (x\, +y) \cdot \, (x^2 \, +\, y^2 \, + \, 30 \, x \, y) 
\nonumber  \\
\hspace{-0.6in}&&\qquad \quad \quad  +2^5 \cdot \,(3\,x^2+3\,y^2\, -62\,x\,y) 
\, \, \, \, \,  
- 2^4 \cdot \,(x+y)\, \, \, \,  \,  +1\,\,\, \, \,  = \,\,\,\, \,  \, 0. 
\end{eqnarray}
This genus-zero curve has the following polynomial parametrization:
\begin{eqnarray}
\hspace{-0.2in}x \, = \, \, {{ (t\, -1)^4} \over {64}}, \qquad \qquad 
y\, = \, \, {{ (t \, +\, 1)^4} \over {64}}.
\end{eqnarray}
In the $\, y= \, x$ limit the  singular variety (\ref{singBat5}) 
gives $\,(1-64\,x) \cdot \,(1\,+16\,x)^2 = \,0$, in agreement with 
the singularities of the 
order-four Calabi-Yau operator (\ref{Batyrev5}). 

\subsection{Last Picard-Fuchs system with two variables}
\label{picardmorebat6}
\vskip .1cm 

A last example  is the two-variables Picard-Fuchs system 
``above'' the order-four Calabi-Yau operator 
(see ODE number 19 in~\cite{TablesCalabi}))
\begin{eqnarray}
\label{Batyrev6}
\hspace{-0.7in}&&529 \, \theta^4 \,\,\, \, \, -23 \,\, x \cdot 
(921\, \theta^4 \, +2046 \,\theta^3 \, 
+1644 \,\theta^2 \,+621 \,\theta \, +92)
\,\nonumber \\
\hspace{-0.7in}&&\quad \quad -\, x^2 \cdot
 (380851\, \theta^4 \, +1328584 \,\theta^3 \,
 +1772673 \,\theta^2 \, +1033528 \,\theta \, +221168) 
 \, \nonumber\\
\hspace{-0.7in}&&\quad \quad 
-\, 2 \, x^3 \cdot
 (475861\, \theta^4 \, +1310172 \,\theta^3 \,
 +1028791 \,\theta^2 \, +208932 \,\theta \, -27232) 
 \,\,\nonumber \\
\hspace{-0.7in}&&\quad \quad -\, 68 \, \, x^4 \cdot
(8873\, \theta^4 \, +14020 \,\theta^3 \,
 +5139 \,\theta^2 \, -1664 \,\theta \, -976)
\, \nonumber\\
\hspace{-0.7in}&&\quad \quad 
+6936 \, \, x^5 \cdot   (3\, \theta \, +\, 4)
 \cdot   (3\, \theta \, +\,2) \cdot
   (\theta \, +\, 1)^2. 
\, 
\end{eqnarray}

The Picard-Fuchs system 
 corresponds to the double series~\cite{Batyrev} 
\begin{eqnarray}
\label{defBat6def}
\hspace{-0.9in}&&\quad \sum_{n=0}^{\infty} \,  \sum_{m=0}^{\infty} \,  
{{(n+m)! \,  \,   (2\, n \, +\, m)! \, \, (2\, m \, +\, n)!
} \over { n!^4 \, \,m!^4 }} 
\cdot \, x^n \, y^m  \, \, \,\, = 
\nonumber  \\
\hspace{-0.9in}&&\quad  \, \,\,\,   \, = \, \, \,  \, 
 \sum_{m=0}^{\infty} \, 
 \, {{(2\, m)!} \over {m!^2}} \, \cdot \, 
_4F_3\Bigl([m\, +1, \,m\, +{{1} \over {2}}, \, 2\,m\, +1, \, {{m\,+\, 1} \over {2}}], 
\, [1, \, 1, \, 1]; \, 4 \, y   \Bigr)
 \cdot \, x^n \nonumber 
\end{eqnarray}
\begin{eqnarray}
\label{defBat6}
\hspace{-0.9in}&&\quad \quad \, \, = \, \, \,  \, 
1 \,\, \, + 2 \cdot \, (x+\, y)\, \, +(6\, x^2+6\, y^2 \, +72\, x\, y) \,  
+ 20 \cdot \, (x\, +y) \cdot \,  ({x}^{2}+ {y}^{2} \, +53\,xy) \, 
\nonumber  \\
\hspace{-0.9in}&&\quad \quad \quad \qquad  \, \,
 +\, 10 \cdot \,
 (1120\,x{y}^{3}+1120\,{x}^{3}y \,  +7\,{x}^{4}+7\,{y}^{4}\, +4860\,{x}^{2}{y}^{2} )
 \,\, \, + \,  \, \cdots 
\end{eqnarray}
Note that all the coefficients of odd orders in $\, x$ and $\, y$ 
factor $(x+y)$.

The  singular variety is the union of 
$\,\,\, x \, y \cdot \, (x+y)  \, = \, 0\,\,$ 
together with the $(x,\,y)$-symmetric  {\em genus-zero} algebraic curve 
which reads:
\begin{eqnarray}
\label{singBat6}
\hspace{-0.6in}&&27 \cdot \, x^2\, y^2\,\cdot \, (y+x)\,\,\,
 -\,[256\, (x^4\, +y^4) \, +304 \, x \, y \, \cdot \,
 (x^2\, + \, y^2) \, +69\, x^2\, y^2]\,\,
\nonumber  \\
\hspace{-0.6in}&&\quad \quad \, +8 \cdot \, (y+x) \cdot \,
 [32\, (x^2\, +y^2)\,   +339\, x\, y]
\nonumber  \\
\hspace{-0.6in}&&\qquad \quad \,
 - \, [96\, (x^2\,+ y^2) \,  - 1261\, x\, y] 
\,\,\,\, 
+16 \cdot \, (y+x)\,\,\, \,-1
\,\,\, \, \, \, = \,\,\,\,\, \,  \, 0. 
\end{eqnarray}
with the simple rational parametrization (see section (\ref{Horn})):
\begin{eqnarray}
\hspace{-0.5in}(x, \, y) \, \, = \, \,\, \,\, 
\Bigl( {{t^4} \over {(t+1) \, (t+2) \, (2 \, t \, +1)^2}},
\, \, \,  {{1} \over {(t+1) \, (t+2)^2 \, (2 \, t \, +1)}}
\Bigr).
\end{eqnarray}
In the $\, y \, = \, \, x$ limit, this singular variety reduces
to  $\, (1 \, -54\,x) \cdot \,(1\,+11\,x \,-x^2)^2\, = \, \,0$
in agreement with the singular points of (\ref{Batyrev6}).

\end{document}